\documentclass[11pt]{article}
\usepackage{amsfonts}
\usepackage{hyperref}
\usepackage{subfigure}
\usepackage{amsmath}
\usepackage{amssymb}
\usepackage{bbold}
\usepackage{geometry}
\usepackage{longtable}
\usepackage{cite}
\usepackage{lscape,graphicx} 
\usepackage{graphics}
\usepackage{color}
\usepackage[numbers,sort&compress]{natbib}

\newcommand\vRto{\mathrel{\stackrel{\makebox[0pt]{\mbox{\normalfont\tiny \text{$v_R$}}}}{\text{$\longrightarrow$}}}}

\newcommand{\trace}[1]{\ensuremath{\mathrm{Tr}#1}}

\begin{document}

{\small
\begin{flushright}
CP3-Origins-2018-026 DNRF90 \\[30pt]
\end{flushright} }

\begin{center}
{\Large \bf  Safe Pati-Salam }\\[20pt]
E.~Molinaro$^{1}$,~F.~Sannino$^{2,3}$,~Z.W. Wang$^{2,4}$\\[10pt]
{\it \small
$^{1}$Department of Physics and Astronomy, University of Aarhus, \\ Ny Munkegade 120, DK-8000 Aarhus C, Denmark\\[4pt]
 $^2$CP$^3$-Origins, University of Southern Denmark,\\
Campusvej 55, DK-5230 Odense M, Denmark\\[4pt]
$^{3}$Danish IAS, University of Southern Denmark, Denmark\\[4pt]
$^{4}$Department of Physics, University of Waterloo, Waterloo, On N2L 3G1, Canada}

\begin{abstract}
We provide an asymptotically safe Pati-Salam embedding of the Standard Model.  Safety is achieved by adding to the theory gauged vector-like fermions and by employing recently developed large number-of-flavor techniques and results. 
 We show that the gauge, scalar quartic and Yukawa couplings achieve an interacting ultraviolet fixed point below the Planck scale. The minimal model is a relevant example of a Standard Model extension in which unification of {\it all} type of couplings occurs because of a dynamical principle, i.e. the presence of an ultraviolet fixed point. This  extension differs from the usual Grand Unified Theories scenario in which only gauge couplings unify and become free  with the remaining couplings left  unsafe. We find renormalization group flow solutions that match the Standard Model couplings values at low energies allowing for realistic safe extensions of the Standard Model. 
\end{abstract}

\end{center}

\section{Introduction}

The recent discovery of  four dimensional asymptotically safe quantum field theories \cite{Litim:2014uca,Litim:2015iea} has opened  the way to safe extensions of the Standard Model, starting with the envision of a {\it safe} rather than {\it free}  QCD \cite{Sannino:2015sel}, to scenarios in which the gauge, the Yukawa and scalar quartic couplings are unified by a dynamical rather than a symmetry principle \cite{Abel:2017ujy,Pelaggi:2017wzr,Abel:2017rwl,Bond:2017wut}.  On the supersymmetric front, exact non perturbative results and constraints were first discussed coherently in \cite{Intriligator:2015xxa}, extending and correcting the results of \cite{Martin:2000cr} while opening the way to  (non)perturbative supersymmetric safety in \cite{Bajc:2016efj,Bond:2017suy,Bajc:2017xwx}, and to the first applications for super GUT model building \cite{Babu:2018tfi,Bajc:2016efj}.  
 Simultaneously there has been much advancement in  our understanding of the nonsupersymmetric dynamics of large number of flavors gauge-Yukawa theories \cite{PalanquesMestre:1983zy,Gracey:1996he,Holdom:2010qs,Pica:2010xq,Shrock:2013cca}. This has led, for example, to enrich the original conformal window \cite{Sannino:2004qp,Dietrich:2006cm}, reviewed in \cite{Sannino:2009za,Pica:2017gcb}, with a novel asymptotically safe region \cite{Antipin:2017ebo}. The discovery led to the upgraded {\it conformal window 2.0} of \cite{Antipin:2017ebo}.
The large $N_f$ dynamics of gauge-fermion theories has been extended to gauge-Yukawa theories starting with the Yukawa sector \cite{Kowalska:2017pkt,Ferreira:1997rc,Ferreira:1997bi} and for the first time to all couplings in \cite{Pelaggi:2017abg,Antipin:2018zdg}.  A gauge-less study appeared in \cite{Alanne:2018ene}. The results  widened the palette of tools and theories at our disposal for novel large $N_f$ safe extensions of the SM \cite{Mann:2017wzh,Abel:2017rwl,Pelaggi:2017abg,McDowall:2018tdg,Ipek:2018sai}.  

We use the acquired  knowledge to construct a novel {\it safe Pati-Salam} extension by adding vector-like fermions and showing that all couplings acquire an UV fixed point at energies that are far from the onset of quantum gravity.  The separation of scales allow us to investigate a condense-matter-like unification of the SM couplings before having to consider the gravitational corrections.  The interplay with gravity has been investigated in several recent works ~\cite{Eichhorn:2018vah,Eichhorn:2017muy,Reichert:2017puo,Eichhorn:2017sok,Eichhorn:2017lry} and it will not be considered here. Differently from the usual Grand Unified scenarios \cite{Georgi:1974sy} in which only the gauge couplings unify because of their embedding into a larger group structure and then they eventually become free, in the present scenario  we have that Yukawa and scalar self couplings are intimately linked because of the safe dynamics with their high energy behavior tamed by the presence of an interacting fixed point.

The paper is organized as follows: In section \ref{Sec2} we review and introduce the Pati-Salam \cite{Pati:1974yy} extension of the SM and  build the minimal vector-like structure able to support a safe scenario. We develop the renormalization group (RG) equations and determine the couplings' evolution in section \ref{Sec3} . Here we analyze and classify the UV fixed point structure of the model. We discuss how to match the SM couplings at low energies in section \ref{Sec4}. We offer our conclusions in section \ref{Sec5}. In appendix~\ref{App1} we summarize the one-loop RG equations for the Pati-Salam model investigated here.

\section{Pati-Salam  extension of the Standard Model}\label{Sec2}
Consider the time-honored Pati-Salam gauge symmetry group $G_\text{PS}$ \cite{Pati:1974yy}
\begin{equation}
	G_\text{PS} = SU(4)\otimes SU(2)_{L} \otimes SU(2)_{R}\,,
\end{equation}
with gauge couplings $g_4$, $g_L$ and $g_R$, respectively. Here the gauge group $SU(4)\supset SU(3)_C \otimes U(1)_{B-L}$, where $SU(3)_C$ denotes the SM color gauge group, and the corresponding gauge couplings are related according to  \begin{eqnarray}
    &&	g_3 =g_4\,,	\quad\quad   g_{B-L} = \sqrt\frac{3}{8} g_3\,.\label{matching_1}
\end{eqnarray}
The gauge fields of $G_\text{PS}$ can be written as follows:
\begin{equation}
\begin{split}
	\hat{W}_{L\mu} & \equiv \; \frac 12 \left( \begin{array}{cc}
						             W^0_{L\mu} & \sqrt{2} W^+_{L\mu} \\
							     \sqrt{2} W^-_{L\mu} & -W^0_{L\mu}
					                      \end{array}\right)\,,
\end{split}
\end{equation}
\begin{equation}
\begin{split}
	\hat{W}_{R\mu} & \equiv \; \frac 12 \left( \begin{array}{cc}
						             W^0_{R\mu} & \sqrt{2} W^+_{R\mu} \\
							     \sqrt{2} W^-_{R\mu} & -W^0_{R\mu}
					                      \end{array}\right)\,,
\end{split}
\end{equation}
\begin{equation}
\begin{split}
	\hat{G}_\mu & \equiv \; \frac 12 \left( \begin{array}{cccc}
						             G_{3\mu} + \frac{G_{8\mu}}{\sqrt 3}+ \frac{B_\mu}{\sqrt 6} & \sqrt 2 G^+_{12\mu} & \sqrt 2 G^+_{13\mu} & \sqrt 2 X^+_{1\mu} \\
							     \sqrt 2 G^-_{12\mu} & -G_{3\mu}+\frac{G_{8\mu}}{\sqrt 3}+ \frac{B_\mu}{\sqrt 6} & \sqrt 2 G^+_{23\mu} & \sqrt 2 X^+_{2\mu} \\
							     \sqrt 2 G^-_{13\mu} & \sqrt 2 G^-_{23\mu} & -\frac{2 G_{8\mu}}{\sqrt 3}+\frac{B_\mu}{\sqrt 6} & \sqrt 2 X^+_{3\mu} \\
							     \sqrt 2 X^-_{1\mu} & \sqrt 2 X^-_{2\mu} & \sqrt 2 X^-_{3\mu} & -\frac{3 B_\mu}{\sqrt 6} 
					                      \end{array}\right)\,.
\end{split}
\end{equation}
In this parametrization, $W^0_{L\mu}$ and $W^\pm_{L\mu}$ correspond to the electroweak (EW) gauge bosons, $G_{3\mu}$, $G_{8\mu}$, $G_{12\mu}^\pm$, $G_{13\mu}^\pm$ and
$G_{23\mu}^\pm$ are the $SU(3)_C$ gluons, $B_\mu$ is the $B-L$ gauge field, and $X_{1\mu}^\pm$, $X_{2\mu}^\pm$ and $X_{3\mu}^\pm$ are leptoquarks. 

The SM quark and lepton fields are unified into the $G_{\rm PS}$ irreducible representations
\begin{eqnarray}
\begin{split}\label{fermionsLR}
	\psi_{L i} &= \left(\begin{array}{cccc} u_L  & u_L & u_L & \nu_L\\ d_L  & d_L & d_L & e_L\end{array}\right)_i \sim  (4,2,1)_i \,, \\ 
	\psi_{R i} &= \left(\begin{array}{cccc} u_R  & u_R & u_R & \nu_R\\ d_R  & d_R & d_R & e_R\end{array}\right)_i \sim (4,1,2)_i \,,  
\end{split}
\end{eqnarray}
where $i=1,2,3$ is a flavor index. 

In order to induce the breaking of $G_{\rm PS}$ to the SM gauge group, we introduce a scalar field $\phi_R$ which transforms as the fermion multiplet $\psi_R$, that is $\phi_R\sim (4,1,2)$:
\begin{equation}
	\phi_R \; = \; \left(\begin{array}{cccc}  \phi_R^u & \phi_R^0 \\ \phi_R^d  & \phi_R^-\end{array} \right)\,,
\end{equation}
where the neutral component $\phi_R^0$ takes a non-zero vev, $v_R\equiv \langle\phi_R^0 \rangle$, such that $G_{\rm PS} \vRto SU(3)_C \otimes SU(2)_L \otimes U(1)_Y$. The hypercharge $Y$ is a linear combination between the diagonal generator of  $SU(2)_R$ and the generator of $B-L$, namely
\begin{equation}
	Y \; = \; 2 \, I_R\,+\,(B-L)\,,
\end{equation}
with ${\rm Tr}\left( I_R^2\right)=1/2$ for the fundamental representation. Then, the EW gauge couplings $g_2$ and $g_Y$  result:
\begin{eqnarray}
	&& g_2 = g_L\,,\quad\quad  g_R = \frac{g_Y}{\sqrt{1-2 g_Y^2/3g_3^2}}\,.\label{matching_2}
\end{eqnarray}

We also introduce an additional (complex) scalar field $\Phi\sim (1,2,2)$, with
\begin{eqnarray}
	\Phi & = & \left(\begin{array}{cc} \phi_1^0 & \phi_2^+ \\ \phi_1^- & \phi_2^0 \end{array} \right) \equiv \left(\begin{array}{cc} \Phi_1 & \Phi_2 \end{array}\right)\,,
\end{eqnarray}
which is responsible of the breaking of the EW symmetry.

\subsection{The Scalar sector}

The general scalar potential of the model defined above is given by:

\begin{eqnarray}
\begin{split}
	V(\Phi, \phi_R)  
	=& -\mu^2_1 \, \trace\left(\Phi^\dagger\Phi\right) + {\rm Re}\left[\mu_{12}^2 \trace\left(\Phi^\dagger\Phi^c\right)\right] -\mu^2_R \trace\left(\phi_R^\dagger\phi_R\right)\\
	&+\lambda_1\,\trace^2\left(\Phi^\dagger\Phi \right)+\rm{Re}\left[\lambda_2\trace^2\left(\Phi^\dagger\Phi^c\right)\right]
	+\rm{Re}\left[\lambda_3\trace\left(\Phi^\dagger\Phi\right)\trace\left(\Phi^\dagger\Phi^c\right)\right]\\
	&+\left(\lambda_4-2\rm{Re}\lambda_2\right)\left\vert\trace\left(\Phi^\dagger\Phi^c\right)\right\vert^2\\
	 			&  +\lambda_{R1} \trace^2\left(\phi_R^\dagger\phi_R \right)
				+ \lambda_{R2}\,\trace\left( \phi_R^\dagger\phi_R \phi_R^\dagger\phi_R \right) \\
			    &+ \lambda_{R\Phi 1}	\trace\left( \phi_R^\dagger\phi_R \right) \trace\left(\Phi^\dagger\Phi\right)  \,+\,\rm{Re}\left[			\lambda_{R\Phi_2}\trace\left(\phi_R\phi_R^\dagger\right)\trace\left(\Phi^\dagger\Phi^c\right) \right]   \\ 
			    & +\lambda_{R\Phi 3} \trace\left( \phi_R^\dagger\phi_R \Phi^\dagger\Phi \right)\,.\label{potential}
\end{split}\label{scalar potential}		
\end{eqnarray}
The quartic couplings $\lambda_{2,3}$ and $\lambda_{R\Phi 2}$, and the dimensional term $\mu_{12}$, carry a non-trivial phase in case CP symmetry is explicitly broken.
 We have also introduced the conjugate field $\Phi^c \equiv \tau_2 \Phi^* \tau_2$, $\tau_2$ being the standard Pauli matrix.

\subsection{The Yukawa sector}

The  most general Yukawa Lagrangian for the matter fields $\psi_{L/R}$ is~\footnote{We consider for simplicity only Yukawa couplings to the third fermion generation and we omit the 
flavor index $i$ in $\psi_{L/R}$, see Eq.~(\ref{fermionsLR}).}
\begin{eqnarray}
\mathcal{L}_{\rm Yuk}^\psi & = & y \, \text{Tr}\left[\overline{\psi_L}\, \Phi\, \psi_R \right] \,+\, y_c \, \text{Tr}\left[\overline{\psi_L}\, \Phi^c\, \psi_R \right]\,+\, \text{h.c.}\,.\label{LYuk1}
\end{eqnarray}
In terms of the SM fermion fields Eq.~(\ref{LYuk1}) reads:
\begin{eqnarray}
\begin{split}
	 \mathcal{L}_{\rm Yuk}^\psi \; =\;  & y \left( \overline{t_L}\, t_R\, \phi_1^0 + \overline{t_L}\, b_R\, \phi_2^+ + \overline{b_L}\, t_R \,\phi_1^- + \overline{b_L}\, b_R\, \phi_2^0 \right.\\
	  &\quad+ \left.\overline{\nu_L} \,\nu_R \,\phi_1^0 + \overline{\nu_L}\, \tau_R \,\phi_2^+ + \overline{\tau_L} \,\nu_R \,\phi_1^- + \overline{\tau_L}\, \tau_R\, \phi_2^0\right)  \\
	 & + y_c \left(  \overline{t_L}\, t_R\, \phi_2^{0*} - \overline{t_L}\, b_R\, \phi_1^+ -  \overline{b_L}\, t_R\, \phi_2^- + \overline{b_L} \,b_R\, \phi_1^{0*} \right. \\
	  & + \left.  \overline{\nu_L}\, \nu_R\, \phi_2^{0*} - \overline{\nu_L}\, \tau_R\, \phi_1^+ - \overline{\tau_L}\, \nu_R\, \phi_2^- + \overline{\tau_L}\, \tau_R \,\phi_1^{0*} \right)\,
	  \;+\; {\rm h.c.}\label{LYuk2}
\end{split}	                 
\end{eqnarray}
Electroweak symmetry breaking is induced by a nonzero vev of $\Phi$, which takes the form:
\begin{equation}\label{vevPhi}
	\langle \Phi \rangle \; = \; \left( \begin{array}{cc} u_1 & 0 \\ 0 & u_2 \end{array} \right)\,,
\end{equation}
with generally $u_1\neq u_2$.
From Eq.~(\ref{LYuk2}) we have the fermion mass spectrum:
\begin{eqnarray}
\begin{split}\label{spectrum1}
	m_t = m_{\nu_\tau} =  (y \sin \beta + y_c \cos\beta) v\,,\\
	m_b = m_\tau = (y \cos\beta + y_c \sin\beta ) v\,,
\end{split}	
\end{eqnarray}
where $v\equiv \sqrt{u_1^2+u_2^2}=174$ GeV and $\tan\beta\equiv u_1/u_2$. In the case of a self-conjugate bi-doublet field $\Phi \equiv \Phi^c$,  one has $u_1=u_2$ in Eq.~(\ref{vevPhi}) and equality between fermion masses is enforced at tree-level, namely
\begin{equation}
	m_t = m_b = m_\tau = m_{\nu_\tau}\,.\label{spectrum2}
\end{equation}
 In order to separate the neutrino and top masses in Eq.~(\ref{spectrum1}) and Eq.~(\ref{spectrum2}) we implement the seesaw mechanism 
\cite{Minkowski:1977sc,Yanagida:1979as,GellMann:1980vs,Mohapatra:1979ia} by adding a new chiral fermion singlet  $N_L\sim (1,1,1)$, which has Yukawa interaction
\begin{equation}
	\mathcal{L}_{\rm Yuk}^N \; = \; - y_\nu \,\overline{N_L} {\rm Tr}\left[ \phi_R^\dagger \,\psi_R \right]\,+\,{\rm h.c.}
\end{equation}
The latter generates a Dirac mass term  $M_R \overline{N_L} \nu_R$, with $M_R\equiv y_\nu v_R$. The resulting Majorana mass term for the neutral fermion fields reads:
\begin{eqnarray}
	\mathcal{L}_{\rm mass}^\nu \;=\; - \frac12 \left(\begin{array}{ccc} \overline{\nu_R^c} & \overline{\nu_R} & \overline{N_R^c} \end{array}\right)
						    \left(\begin{array}{ccc} 0 & m_t & 0 \\ 
						    				       m_t & 0 & M_R \\
										       0 & M_R & 0  \end{array}\right)
						    \left(\begin{array}{c} \nu_L \\ \nu_L^c \\ N_L \end{array}\right)\, + \,  {\rm h.c.} 		       
\end{eqnarray}
with $\nu^c_{L/R}\equiv C \overline{\nu}_{R/L}^T$ and $N^c_R \equiv C\overline{N_L}^T$. The mass spectrum consists of one massless neutrino 
\begin{equation}
	\nu_{\tau L} = -\cos\theta \nu_{L} + \sin\theta N_L\,,
\end{equation}
with $\tan\theta = m_t/M_R$, and one Dirac neutrino $N_D$ with mass $m_{D}=\sqrt{m_t^2+M_R^2}$ and chiral components:
\begin{eqnarray}
\begin{split}
	P_L \,N_D \;=\; & \nu_{L} \sin\theta + N_L \cos\theta \,,\\
	P_R\, N_D \;=\; & \nu_R \,.
\end{split}
\end{eqnarray}
By adding a Majorana mass term for the singlet fermion $N_L$
\begin{equation}
	\mathcal{L}_{\rm mass}^N = -\frac 12 M_N \overline{N^c_R} \, N_L\,+\, {\rm h.c.}
\end{equation}
the total lepton number is explicitly broken and the spectrum consists of three massive neutrinos. Taking $M_N\ll m_t, M_R$, the mass eigenstates result in one light active Majorana neutrino $\nu_{\tau}$ with mass
\begin{equation}
	m_{\nu_\tau} =  M_N \frac{m_t^2}{m_D^2}
\end{equation}
and two quasi-degenerate heavy Majorana neutrinos $N_{1,2}$ with opposite CP parities and masses
\begin{equation}
	M_{1,2} = m_D \pm \frac{M_N}{2} \frac{M_R^2}{m_D^2}\,.
\end{equation}
Threshold corrections may induce a sizable  mass splitting between $m_\tau$ and $m_b$ in Eq.~(\ref{spectrum1}) and Eq.~(\ref{spectrum2}), which depends on the $G_{\rm PS}$ breaking scale $v_R$, see Ref.~\cite{Volkas:1995yn}. 

\subsection{The minimal model}

In the simplest scenario where the field $\Phi$ is self-conjugate, the fermion spectrum is  degenerate, see Eq.~(\ref{spectrum2}), and 
the scalar potential in Eq.~(\ref{potential}) consists of the quartic couplings $\lambda_1$, $\lambda_{R1,2}$ and $\lambda_{R\Phi1,3}$.
 As discussed above, by adding a new chiral fermion $N_L$, which is a singlet under $G_{\rm PS}$, it is possible to induce a hierarchy between the top quark and neutrino masses via the seesaw
 mechanism, such that the correct light neutrino mass scale can be accommodated. Here, we further extend the
matter content of the theory with a new vector-like fermion $F\sim (10, 1, 1)$ with mass $M_F$ and Yukawa interactions:
\begin{eqnarray}\label{YukF}
	\mathcal{L}_{\rm Yuk}^F & = & y_F\, {\rm Tr}\left( \overline{F_L} \, \phi_R^T \,i \tau_2 \,\psi_R \right) \,+\, {\rm h.c.}	\label{Yuk_F}
\end{eqnarray}
In terms of the $SU(3)_C$ representations, the field $F$ can be decomposed as
\begin{equation}
	F \;= \; \left(\begin{array}{cc} S & B\sqrt{2} \\ B^T\sqrt{2} & E \end{array}\right)\,,
\end{equation}
where $S$, $B$ and $E$ denote a  color sextet, triplet and singlet, respectively. 
Then, from Eq.~(\ref{YukF}) the fields $B$ and $E$ mix with the
right-handed components of $\psi_R$, $b_R$ and $\tau_R$, respectively,  giving the overall Dirac mass terms:
\begin{eqnarray}
	\mathcal{L}_{\rm mass}^b & = & \left( \begin{array}{cc} \overline{b_L} & \overline{B_L} \end{array} \right) 
							 \left( \begin{array}{cc} m_t & 0 \\ m_B & M_F \end{array} \right)
							 \left( \begin{array}{c} b_R \\ B_R \end{array}  \right)\,+\, {\rm h.c.}\,, \\
	\mathcal{L}_{\rm mass}^\tau & = & \left( \begin{array}{cc} \overline{\tau_L} & \overline{E_L} \end{array} \right) 
							     \left( \begin{array}{cc} m_t & 0 \\ \sqrt{2}\,m_B & M_F \end{array} \right)
							     \left( \begin{array}{c} \tau_R \\ E_R \end{array}  \right)\,+\, {\rm h.c.}\,, 
\end{eqnarray}
with $m_B\equiv y_F\, v_R/\sqrt{2}$. As a result of this mixing, the top quark becomes naturally heavier than the other SM fermions. 
In fact, in the limit $m_B\gg m_t, \, M_F$, the $b$ quark and $\tau$ charged lepton masses satisfy the tree-level relation:
\begin{equation}
m_b\; =\; \sqrt{2}\, m_\tau \; \approx\; \frac{M_F\,m_t}{\sqrt{2}\, m_B}\,.\label{bottom_tau_mass}
\end{equation}
Analogously, we have a new vector-like  quark, $\hat B$, and a new vector-like lepton, $\hat E$, with corresponding masses $M_B$ and $M_E$, 
which satisfy the tree-level relation:
\begin{equation}
	M_B \; = \; M_E/\sqrt{2} \;\approx\; m_B\,.	
\end{equation}

\section{Renormalization group analysis}\label{Sec3}
In this section, we perform the RG analysis of the Pati-Salam extension of the SM introduced above and discuss the relevant phenomenological implications. 
The gauge, Yukawa and scalar couplings in the minimal and extended realizations  are listed in Tab.~\ref{couplings}. The corresponding RG equations at one loop order are reported in appendix~\ref{RG_all}.
\begin{table}[t!]
\centering
  \begin{tabular}{|| l | l | l ||}
    	\hline
Gauge Couplings & Yukawa Couplings & Scalar Couplings \\ \hline
$SU(4):\,g_4$ & $\psi_{L/R}: \,y,\,y_c$ & $\phi_R:\,\lambda_{R1},\,\lambda_{R2}$\\\hline
$SU(2)_L:\,g_L$ & $N_L:\,y_{\nu}$ & portal: $\lambda_{R\Phi_1},\,\lambda_{R\Phi_2},\,\lambda_{R\Phi_3}$\\ \hline
$SU(2)_R:\,g_R$ & $F :\,y_F$ & $\Phi:\,\lambda_1,\,\lambda_2,\,\lambda_3,\,\lambda_4$\\ \hline
\end{tabular}
\caption{\small Gauge, Yukawa and scalar quartic couplings of the Pati-Salam model.}
\label{couplings}
\end{table}

\mathversion{bold}
\subsection{Large-$N$ beta function}
\mathversion{normal}

\begin{figure}[t]
\centering
\subfigure[]{
\label{bubble_1}
\begin{minipage}{6cm}
\centering
\includegraphics[width=1\columnwidth]{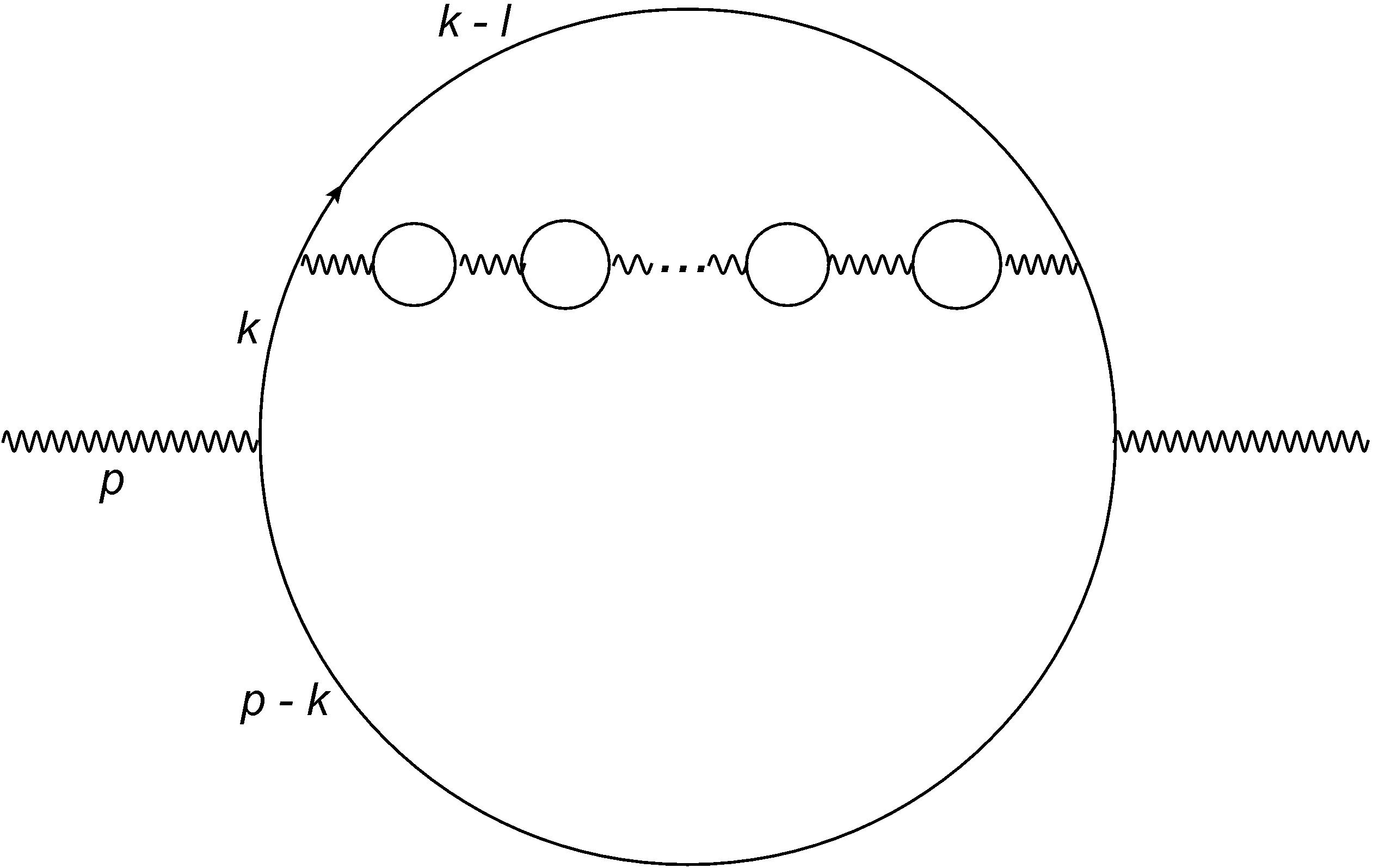}\hspace{0.05\columnwidth}
\end{minipage}
}
\subfigure[]{
\label{bubble_2}
\begin{minipage}{6cm}
\centering
\includegraphics[width=1\columnwidth]{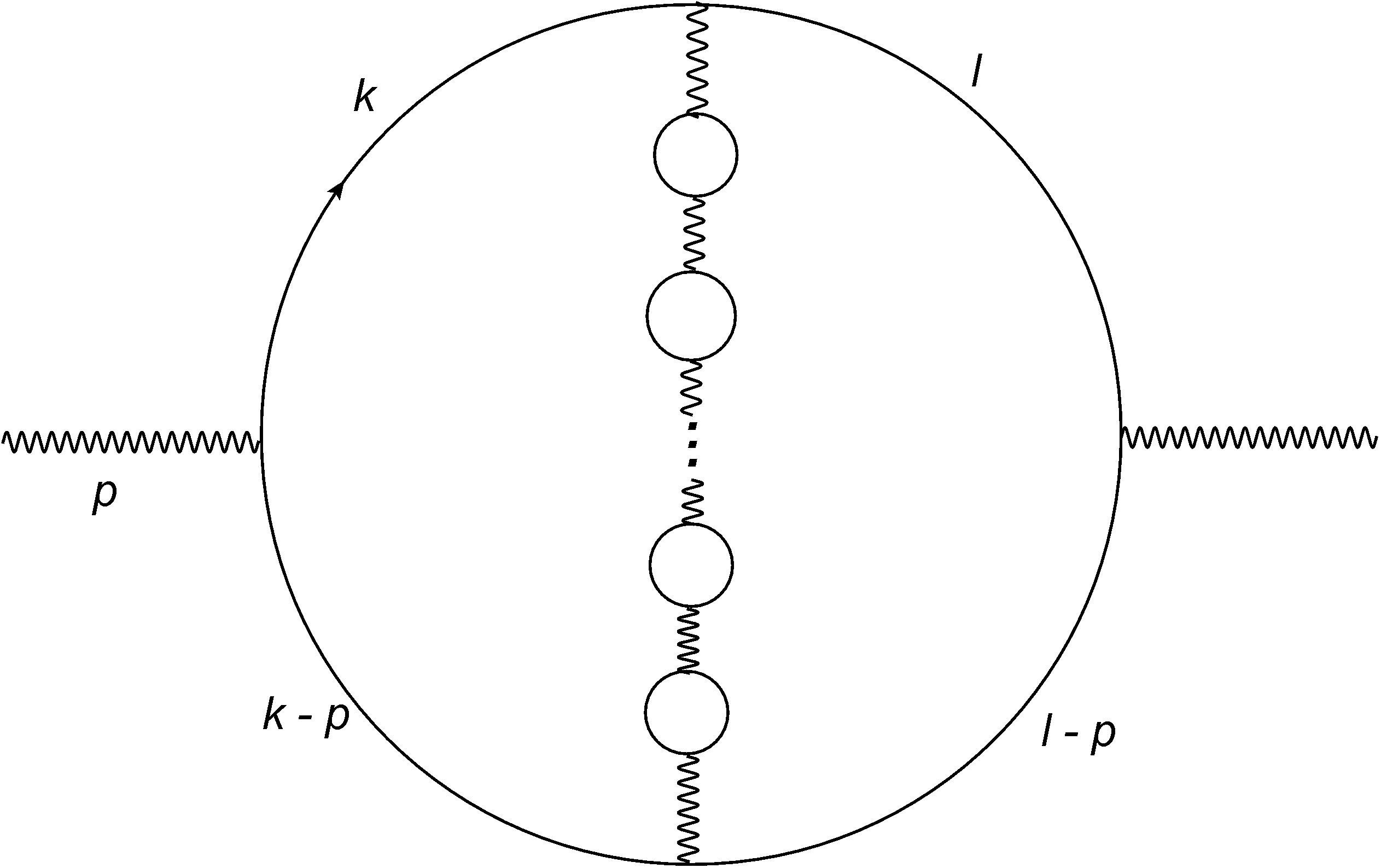}\hspace{0.05\columnwidth}
\end{minipage}
}
\subfigure[]{
\label{bubble_3}
\begin{minipage}{6cm}
\centering
\includegraphics[width=1\columnwidth]{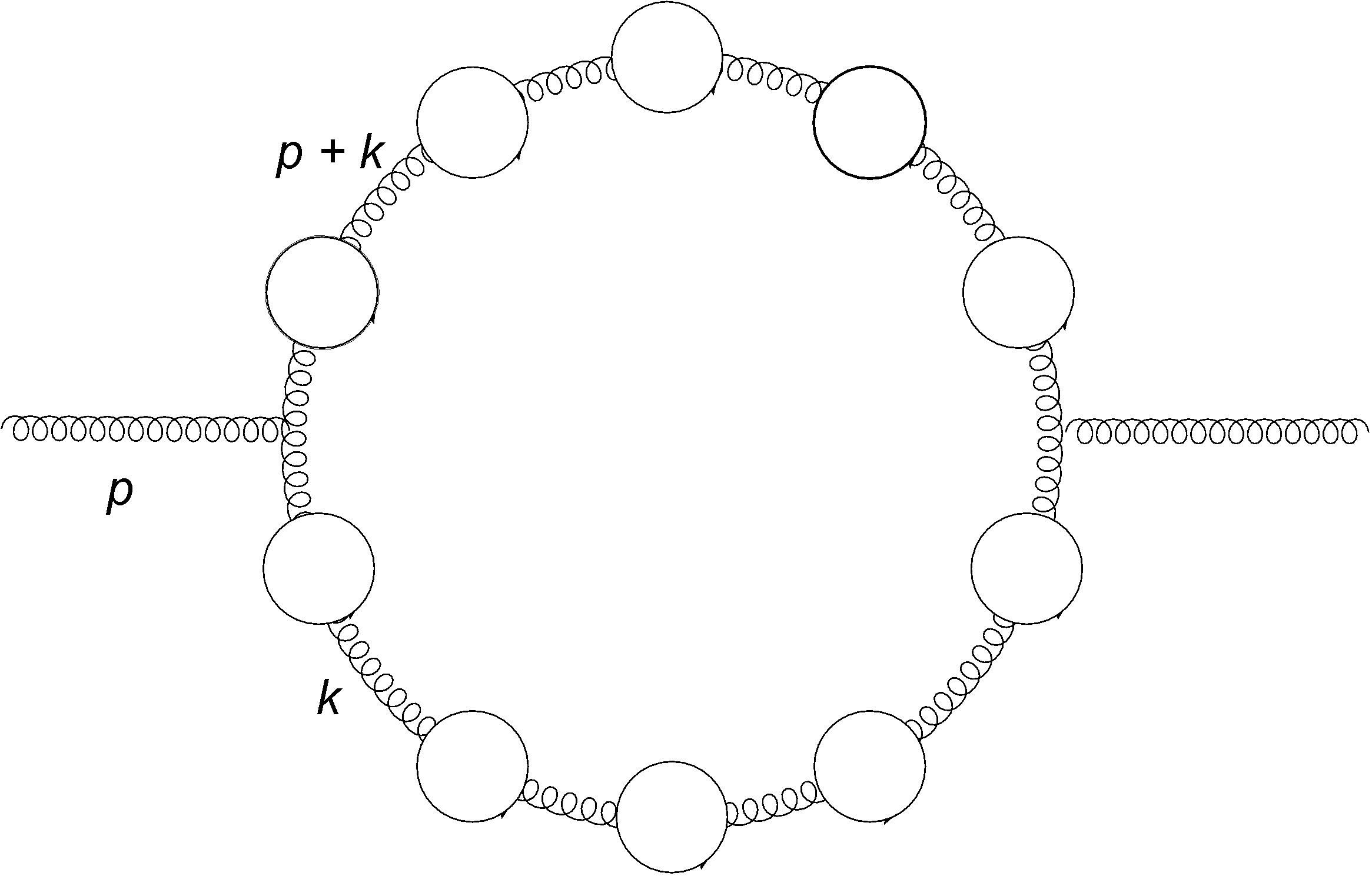}\hspace{0.05\columnwidth}
\end{minipage}
}
\subfigure[]{
\label{gauge_cancel}
\begin{minipage}{6cm}
\centering
\includegraphics[width=1\columnwidth]{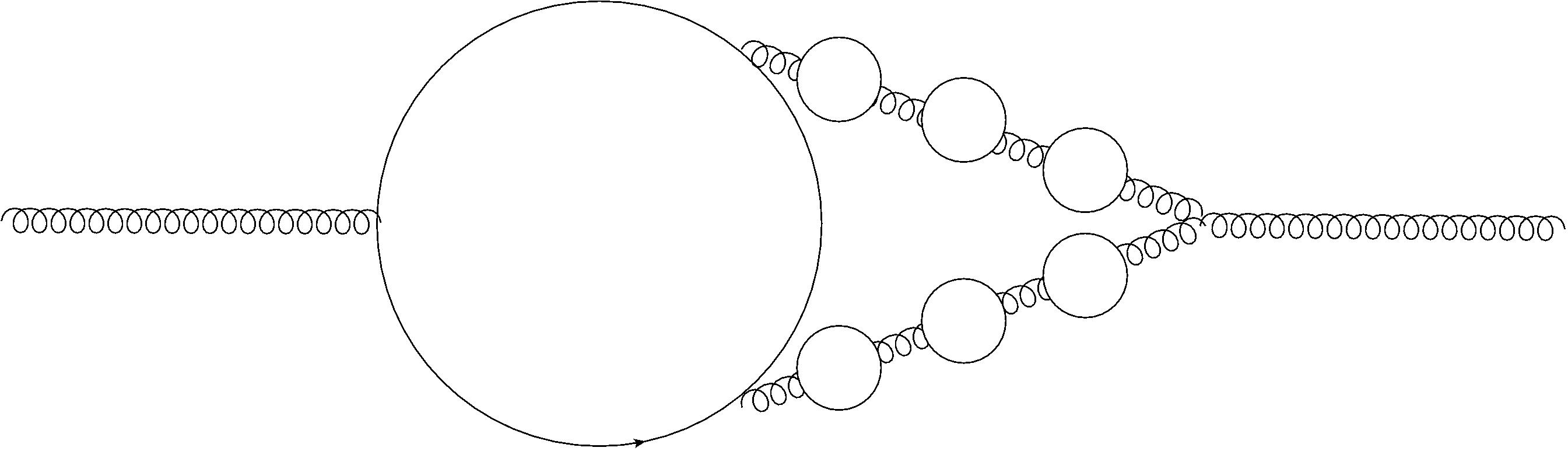}\hspace{0.05\columnwidth}
\end{minipage}
}
\caption{Feynman diagrams for gauge field renormalization at order $1/N_F$. Diagrams (a) and (b) are present in both the Abelian and non-Abelian 2-point functions, while (c) and (d) only exist in the non-Abelian theory. }
\label{bubble diagram}
\end{figure}

In order to ensure asymptotic safety in the UV for all the system in Tab.~\ref{couplings}, we employ the $1/N_F$ expansion approach developed in \cite{PalanquesMestre:1983zy,Gracey:1996he,Holdom:2010qs,Pica:2010mt}, first applied to the whole SM in \cite{Mann:2017wzh}. 
More specifically, we introduce $N_F\gg 1$ vector-like fermions, which transform non-trivially under $G_{\rm PS}$. 
In this framework, the RG equations receive a  contribution at leading order in the $1/N_F$ expansion of the relevant Feynman diagrams, which are resumed as shown 
in Fig~\ref{bubble diagram} (only gauge coupling cases are shown). This non-perturbative effect induces an interacting fixed point for both the Abelian and non-Abelian gauge interactions of the SM \cite{Mann:2017wzh}.
The fixed point is guaranteed by the pole structure occurred in the expressions of the summation \cite{Gracey:1996he, Holdom:2010qs}.

In the present scenario, we consider three sets of vector-like fermions charged under $G_{\rm PS}$,  
with the following charge assignment:
\begin{equation}
N_{F_{4}}\left(4,1,1\right)\oplus N_{F_{2L}}\left(1,3,1\right)\oplus N_{F_{2R}}\left(1,1,2\right)\,,
\end{equation}
where the $N_{F_{2L}}$ vector-like fermions are chosen in the adjoint representation of $SU(2)_L$ to avoid fractional electrical charges. We have also chosen each set of vector-like fermions to have non-trivial charges only under one simple gauge group to avoid the extra contributions in the summation of semi-simple group.

\mathversion{bold}
\subsection{Large-$N$ gauge beta function and gauge coupling unification}
\mathversion{normal}

To the leading $1/N_F$ order for each set, the higher order (ho) contributions (i.e.~the bubble diagrams in Fig.~\ref{bubble diagram}) to the RG functions of the gauge couplings are calculated in  \cite{Holdom:2010qs}, while for the abelian case they were first computed in \cite{PalanquesMestre:1983zy}. Here we list a short summary of the results. The higher order contributions are give by:
\begin{equation}
\beta^{{\rm ho}}_{ i}=\frac{2A_i\alpha_i}{3}\frac{H_{1_i}(A_i)}{N_{F_i}},\quad\alpha_i\equiv\frac{g_i^2}{\left(4\pi\right)^2}~~\left(i=2L,\,2R,\,4\right)\,,\label{higher order contribution}
\end{equation}
with the functions $H_{1i}$ and the t'Hooft couplings $A_i$ given by
\begin{equation}
\begin{split}
A_i&=4\alpha_iT_RN_{F_i}\\
H_{1_i}&=\frac{-11}{2}N_{ci}+\int_0^{A_i/3}I_1(x)I_2(x)dx\qquad\left(N_{ci}=2,4\right)\\
I_1(x)&=\frac{\left(1+x\right)\left(2x-1\right)^2\left(2x-3\right)^2\sin\left(\pi x\right)^3}{\left(x-2\right)\pi^3}\times\left(\Gamma\left(x-1\right)^2\Gamma\left(-2x\right)\right)\\
I_2(x)&=\frac{N_{ci}^2-1}{N_{ci}}+\frac{\left(20-43x+32x^2-14x^3+4x^4\right)}{2\left(2x-1\right)\left(2x-3\right)\left(1-x^2\right)}N_{ci}\,.
\end{split}
\end{equation}
The Dynkin indices are $T_R=1/2~(N_{ci})$ for the fundamental (adjoint) representation.
The RG functions of the gauge couplings (see Appendix \ref{RG_all}) including the contributions of bubble diagrams resummation are listed below:

\begin{equation}
\begin{split}
\beta_{\alpha_{2L}}^{tot}&=\frac{d\alpha_{2L}}{d\log\mu}=\beta_{\alpha_{2L}}^{1loop}+\beta_{\alpha_{2L}}^{\rm{ho}}=-6\alpha_{2L}^2+\frac{2A_{2L}\alpha_{2L}}{3}\frac{H_{1_{2L}}\left(A_{2L}\right)}{N_{F_{2L}}}\\
\beta_{\alpha_{2R}}^{tot}&=\frac{d\alpha_{2R}}{d\log\mu}=\beta_{\alpha_{2R}}^{1loop}+\beta_{\alpha_{2R}}^{\rm{ho}}=-\frac{14}{3}\alpha_{2R}^2+\frac{2A_{2R}\alpha_{2R}}{3}\frac{H_{1_{2R}}\left(A_{2R}\right)}{N_{F_{2R}}}\\
\beta_{\alpha_{4}}^{tot}=&\frac{d\alpha_{4}}{d\log\mu}=\beta_{\alpha_{4}}^{1loop}+\beta_{\alpha_{4}}^{\rm{ho}}=-18\alpha_{4}^2+\frac{2A_{4}\alpha_{4}}{3}\frac{H_{1_4}\left(A_{4}\right)}{N_{F_{4}}}\,,\\
\end{split}
\label{Gauge couplings RG Bubble}
\end{equation}
where the $\beta_{\alpha_{2L}}^{1loop},\,\beta_{\alpha_{2R}}^{1loop},\,\beta_{\alpha_{4}}^{1loop}$ are denoted as the original one loop RG beta functions of the three gauge couplings without bubble diagram contributions while $\beta_{\alpha_{2L}}^{tot},\,\beta_{\alpha_{2R}}^{tot},\,\beta_{\alpha_{4}}^{tot}$ are the total RG beta functions including the higher order bubble diagram contributions up to $1/N_F$ order. The reason that only one loop RG beta functions of the gauge couplings are used will be clear later on.

\begin{figure}[t!]
\centering
\includegraphics[width=0.6\columnwidth]{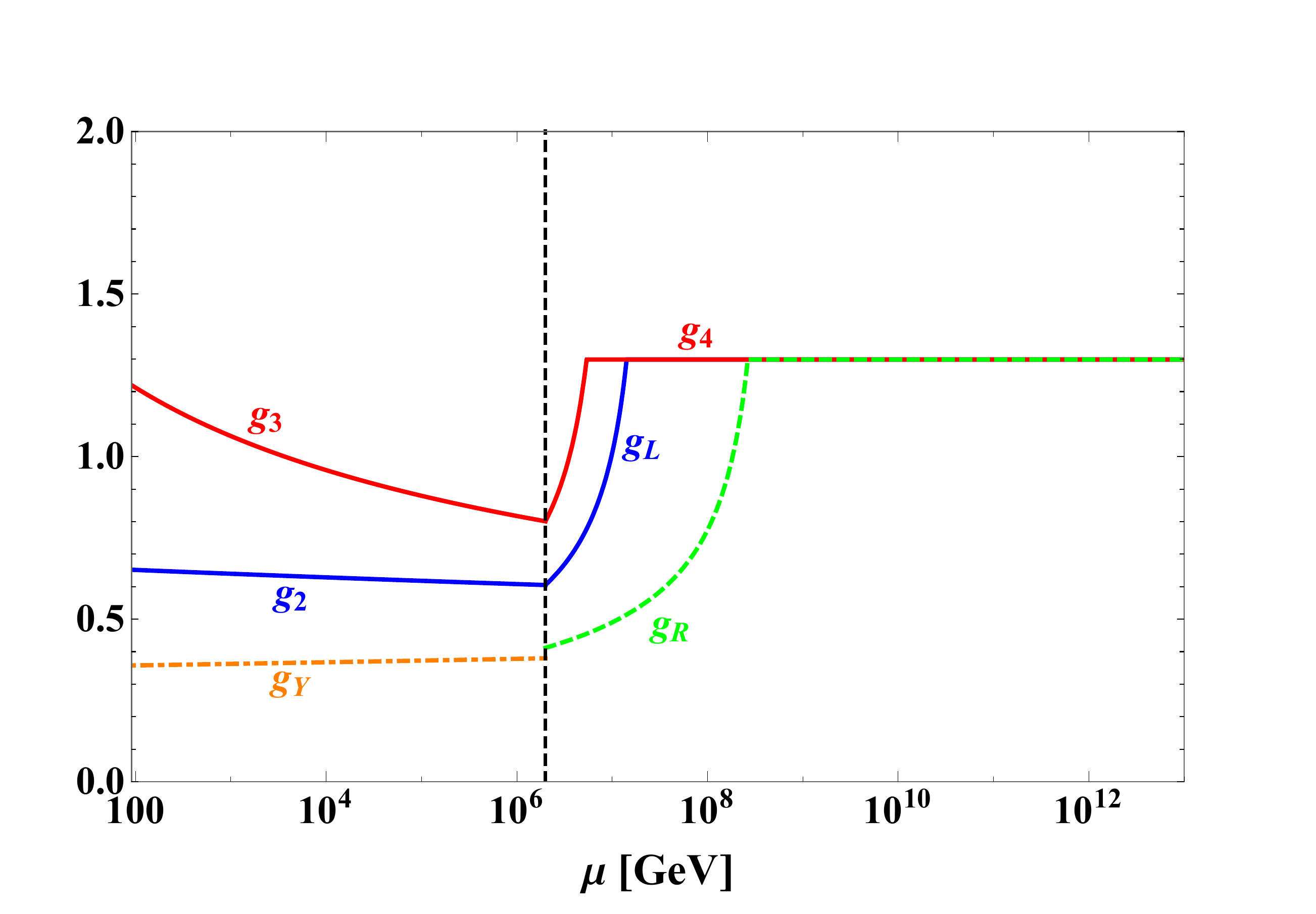}\hspace{0.06\columnwidth}
\caption{\small We show a sample case of gauge unification, where we have chosen $N_{F_{2L}}=35,\,N_{F_{2R}}=N_{F_{4}}=140$. The dashed line represents the Pati-Salam symmetry breaking scale at $2000\,\rm{TeV}$ where all the vector-like fermions are introduced. The three couplings $g_Y,\,g_2,\,g_3$ at the left hand side of the dashed line are determined by the running of the SM gauge couplings.}
\label{unification}
\end{figure}

Thus, the UV fixed point for the gauge coupling sub-system $\left(g_4,\,g_L,\,g_R\right)$ is guaranteed by the pole structure in the bubble diagram summation. 
For all the non-abelian gauge groups, the pole in the function $H_{1_i}$, and thus the UV fixed point of the non-abelian gauge couplings, always occurs at $A_i=3$. 
In particular, if one chooses the vector-like fermion representation with $A_{2L}=A_{2R}=A_{4}$, 
gauge coupling unification is guaranteed. This is shown in Fig.~\ref{unification}, where we set $N_{F_{2L}}=35$ and $N_{F_{2R}}=N_{F_{4}}=140$. The IR initial conditions of $g_L$, $g_R$ and $g_4$ are obtained by using the matching conditions of Eq.~\eqref{matching_1} and Eq.~\eqref{matching_2} and the SM couplings are running from the EW scale to the Pati-Salam symmetry breaking scale. For simplicity, we have assumed all the vector-like fermions were introduced at the Pati-Salam symmetry breaking scale $v_R$. The latter is most strongly constrained by the kaon decay $K_L\rightarrow\mu^{\pm}e^{\mp}$ (see e.g.~\cite{Volkas:1995yn,Valencia:1994cj}). Using the current upper limit ${\rm Br}\left(K_L\rightarrow\mu^{\pm}e^{\mp}\right)<4.7\times10^{-12}$ provided in \cite{Ambrose:1998us},
we obtain the lower limit $v_R\gtrsim 2000$ TeV (see also e.g.~\cite{Parida:2014dba}). In order to make closer connection to low energy phenomenology, in this work we choose the Pati-Salam symmetry breaking scale exactly at $2000\,\rm{TeV}$.

\mathversion{bold}
\subsection{Large-$N$ Yukawa and quartic beta function}
\mathversion{normal}

In the previous section, we have only considered the bubble diagram contributions in the gauge couplings subsystem. However, the bubble diagrams can  directly contribute also  to the quartic and Yukawa beta functions (see e.g.~\cite{Antipin:2018zdg,Kowalska:2017pkt}). In the following, we provide a brief review of the procedure following \cite{Antipin:2018zdg}.

The bubble diagram contributions to known 1-loop beta functions of quartic and Yukawa couplings can be obtained by employing the following recipe. The Yukawa beta function at large number of fermions can be written in the compact form
\begin{equation}
\begin{split}
\beta_y&=c_1y^3 + y \sum_{\alpha}c_\alpha g^2_\alpha I_y\left(A_\alpha\right),\quad\rm{with}\\\label{eq-simplifiedyukawa}
I_y\left(A_\alpha\right)&=H_\phi\left(0,\tfrac{2}{3}A_\alpha\right)\left(1+A_\alpha\frac{C_2\left(R_\phi^\alpha\right)}{6\left(C_2\left(R_{\chi}^\alpha\right)+C_2\left(R_{\xi}^\alpha\right)\right)}\right)\\
H_\phi(x) &=H_0(x)= \dfrac{(1 - \tfrac{x}{3}) \Gamma(4-x)}{3 \Gamma^2(2 - \tfrac{x}{2}) \Gamma(3 - \tfrac{x}{2}) \Gamma(1 + \tfrac{x}{2})}
\end{split}
\end{equation}
containing information about the resumed fermion bubbles and $c_1,\,c_\alpha$ are the standard 1-loop coefficients for the Yukawa beta function while $C_2(R_\phi^\alpha),\,C_2(R_{\chi}^\alpha),\,C_2(R_{\xi}^\alpha)$ are the Casimir operators of the corresponding scalar and fermion fields. Thus, when $c_1,\,c_\alpha$ are known, the full Yukawa beta function including the bubble diagram contributions can be obtained.
Similarly, for the quartic coupling we write
\begin{equation}
\beta_\lambda=c_1\lambda^2+\lambda \sum_{\alpha}c_\alpha \,g^2_\alpha\,I_{\lambda g^2}\left(A_\alpha\right)+\sum_{\alpha} c'_\alpha \,g_\alpha^4\,I_{g^4} \left(A_\alpha\right) +\sum_{\alpha < \beta}c_{\alpha\beta}\, g_\alpha^2 g_\beta^2 \,I^{tot}_{g_1^2g_2^2}\left(A_\alpha,\,A_\beta\right)\,,\label{quartic_bubble_beta}
\end{equation}
with $c_1,\,c_\alpha,\,c'_\alpha,\,c_{\alpha\beta}$  the known 1-loop coefficients
for the quartic beta function and the resumed fermion bubbles appear via 
\begin{equation}
\begin{split}
I_{\lambda g^2}\left(K_\alpha\right) &=H_\phi\left(0,\tfrac{2}{3}A_\alpha\right)\\
I_{g^4}\left(K_\alpha\right)&=H_\lambda\left(1,\tfrac{2}{3}A_\alpha\right)+A_\alpha\frac{dH_\lambda\left(1,\tfrac{2}{3}A_\alpha\right)}{dA_\alpha}\\
I_{g_1^2g_2^2}^{tot}\left(A_\alpha,\,A_\beta\right)&=\frac{1}{3}\left[I_{g_1^2g_2^2}\left(A_\alpha,\,0\right)+I_{g_1^2g_2^2}\left(0,\,A_\beta\right)+I_{g_1^2g_2^2}\left(A_\alpha,\,A_\beta\right)\right]\\
I_{g_1^2g_2^2}\left(A_\alpha,\,A_\beta\right)&=\frac{1}{A_\alpha-A_\beta}\left[A_\alpha H_\lambda\left(1,\tfrac{2}{3}A_\alpha\right)-A_\beta H_\lambda\left(1,\tfrac{2}{3}A_\beta\right)\right],\quad\rm{where}\\
H_\lambda(1,x) &= (1-\tfrac{x}{4}) H_0(x)=\dfrac{ (1-\tfrac{x}{4})(1 - \tfrac{x}{3}) \Gamma(4-x)}{3 \Gamma^2(2 - \tfrac{x}{2}) \Gamma(3 - \tfrac{x}{2}) \Gamma(1 + \tfrac{x}{2})}\,.
\end{split}
\end{equation} 
Thus we have now  the full quartic beta function including the bubble diagram contributions when $c_1,\,c_\alpha,\,c_\alpha',\,c_{\alpha\beta}$ are known. Following the above recipe, the bubble diagram improved Yukawa beta function $\beta_y$, for example, can be written as
\begin{equation}
(4 \pi )^2\beta_y=\left(-\frac{9}{4} g_L^2 I_y\left(A_L\right)-\frac{45 }{4} g_4^2 I_y\left(A_4\right)-\frac{9}{4} g_R^2 I_y\left(A_R\right)+20 y_c^2+y_{\nu }^2\right)y+12 y^3\,.
\end{equation}
The bubble diagram improved quartic beta function $\beta_{\lambda_{R1}}$ reads
\begin{equation}
\begin{split}
(4 \pi )^2\beta_{\lambda_{R1}}&=192 \lambda _{R1}^2+\lambda _{R1} \left(-\frac{45}{2} g_4^2 I_{\lambda g^2}\left(A_4\right)-9 g_R^2 I_{\lambda g^2}\left(A_R\right)+192 \lambda _{R2}+8 y_{\nu }^2\right)\\
&+\frac{27}{32} g_4^2 g_R^2 \times\frac{1}{3}\left(I_{g_1^2g_2^2}\left(A_4,A_R\right)+I_{g_1^2g_2^2}\left(0,A_R\right)+I_{g_1^2g_2^2}\left(A_4,0\right)\right)\\
&+\frac{27}{128} g_4^4 I_{g^4}\left(A_4\right)+\frac{9}{32} g_R^4 I_{g^4}\left(A_R\right)+48 \lambda _{R2}^2+16 \lambda _{R\Phi1}^2
-2 y_{\nu }^4\,.
\end{split}
\end{equation}

\subsection{UV fixed point solutions in the gauge-Yukawa-quartic system}
To prove the existence of a fixed point of the whole system in Tab.~\ref{couplings}, we are entitled to assume the gauge couplings at the UV fixed point as background values (i.e. constants in the RG functions of other couplings). This is so because at the UV fixed point they only depend on the choice of $N_F$. By using the one loop RG functions in appendix \ref{RG_all} augmented with the large-$N$ corrections (i.e.~Eq.~\eqref{eq-simplifiedyukawa} and Eq.~\eqref{quartic_bubble_beta}), we can now set $\{\beta_i=0\}$  where $i$ denotes all the Yukawa and scalar couplings presented in  Tab.~\ref{couplings}. Our investigation and beta functions are consistent with the large-$N$ limit, computations and results established in \cite{Mann:2017wzh,Antipin:2018zdg}.

We impose CP invariance, that implies: ${\rm Im}\left(\lambda_2\right)={\rm Re}\left(\lambda_3\right)={\rm Re}\left(\lambda_{R\Phi2}\right)=0$.  This symmetry requires $y=\pm y_c$, leading to top and bottom mass degeneracy, which is lifted when including the new vector-like fermion $F\sim\left(10,1,1\right)$ (see Eq.~\eqref{Yuk_F}).  We have also checked that, when breaking the CP symmetry safety is lost, because the overall RG system is over-constrained.

The analysis unveils several UV candidate fixed points for different choices of $N_F$. For example, for $N_{F_{2L}}=40,\,N_{F_{2R}}=150,\,N_{F_4}=200$, we discover  30 sets of UV candidate fixed point.  However the scalar potential is unbounded for several candidates UV fixed points. We therefore require the following vacuum stability conditions (see e.g.~\cite{Holthausen:2009uc}) to be satisfied:
\begin{equation}
\lambda_{R1}+\lambda_{R2}>0\,\qquad\lambda_1-\lambda_2+\lambda_4>0,\qquad\lambda_1>0\label{vacuum_stability}\,.
\end{equation} 
These conditions are quite constraining, reducing to 5 the original set of 30 UV fixed point candidates.

Consider the same value of the number of vector-like fermions discussed above (i.e.~$N_{F_{2L}}=40,\,N_{F_{2R}}=150,\,N_{F_4}=200$). We now select one sample UV fixed point solutions summarized in Tab.~\ref{shifting UV fixed point_1}. 
The solutions listed in Tab.~\ref{shifting UV fixed point_1} satisfy the vacuum stability condition Eq.~\eqref{vacuum_stability}. For a different sample value of the number of vector-like fermions ($N_{F_{2L}}=40,\,N_{F_{2R}}=80,\,N_{F_4}=100$), we also find a set of UV fixed point solutions which satisfy the vacuum stability conditions 
(see Tab.~\ref{shifting UV fixed point_2}). 

\begin{table}[t!]
\centering
  \begin{tabular}{|| l | l | l | l | l | l | l | l | l | l | l | l ||}
    	\hline
  $\lambda_1$ & $\lambda_2$ & $\lambda_3$ & $\lambda_4$ & $\lambda_{R\Phi_1}$ & $\lambda_{R\Phi_{2,3}}$ & $\lambda_{R1}$ & $\lambda_{R2}$ & $y$ & $y_c$ & $y_\nu$ & $y_F$ \\ \hline
   0.12 & 0.05 & 0 & 0.13 & 0.02 &  0 & 0.13 & -0.01 & 0.78 & 0.78 & 0.84 & 0 \\ \hline
\end{tabular}
\caption{\small This table summarizes the sample UV fixed point solution with sample value ($N_{F_{2L}}=40,\,N_{F_{2R}}=150,\,N_{F_4}=200$)
 involving the bubble diagram contributions in the Yukawa and quartic RG beta functions.}
\label{shifting UV fixed point_1}
\end{table}

So far $y_F$ was asymptotically free (see Tab.~\ref{shifting UV fixed point_1} and Tab.~\ref{shifting UV fixed point_2},) and we now exhibit the case in which $y_F\neq0$ in the UV.  This case is shown in Tab.~\ref{shifting UV fixed point_3} in which we have a UV safe solution for $y_F$ for ($N_{F_{2L}}=40,\,N_{F_{2R}}=130,\,N_{F_4}=130$).
Interestingly this solution owes its existence to the bubble diagram contributions for the Yukawa and quartic RG beta functions. Thus, the large-$N$ contributions for the Yukawa and quartic couplings add novel safe possibilities in which all Yukawa couplings are safe.

\begin{table}[t!]
\centering
  \begin{tabular}{|| l | l | l | l | l | l | l | l | l | l | l | l ||}
    	\hline
   $\lambda_1$ & $\lambda_2$ & $\lambda_3$ & $\lambda_4$ & $\lambda_{R\Phi_1}$ & $\lambda_{R\Phi_{2,3}}$ & $\lambda_{R1}$ & $\lambda_{R2}$ & $y$ & $y_c$ & $y_\nu$ & $y_F$ \\ \hline
   0.21 & 0.07 & 0 & 0.24 & 0.03 & 0 & 0.27 & -0.02 & 1.05 & 1.05 & 1.19 & 0\\ \hline
\end{tabular}
\caption{\small This table summarizes the UV fixed point solution for ($N_{F_{2L}}=40,\,N_{F_{2R}}=80,\,N_{F_4}=100$) 
involving the bubble diagram contributions in the Yukawa and quartic RG beta functions.}
\label{shifting UV fixed point_2}
\end{table}

\begin{table}[t!]
\centering
  \begin{tabular}{|| l | l | l | l | l | l | l | l | l | l | l | l ||}
    	\hline
   	$\lambda_1$ & $\lambda_2$ & $\lambda_3$ & $\lambda_4$ & $\lambda_{R\Phi_1}$ & $\lambda_{R\Phi_{2,3}}$ & $\lambda_{R1}$ & $\lambda_{R2}$ & $y$ & $y_c$ & $y_\nu$ & $y_F$ \\ \hline
   	0.05 & 0.02 & 0 & 0.01 & 0.04 & 0 & 0.02 & 0.08 & 0.24 & 0.24 & 0.57 & 0.74\\ \hline
\end{tabular}
\caption{\small This table summarizes the  UV fixed point solution for ($N_{F_{2L}}=40,\,N_{F_{2R}}=130,\,N_{F_4}=130$) 
involving the bubble diagram contributions in the Yukawa and quartic RG beta functions. 
}
\label{shifting UV fixed point_3}
\end{table}

We now determine which fixed point is relevant/irrelevant (UV repulsive/attractive) following the convention according to which the RG flows towards the IR.   
The results are summarized in Tab.~\ref{classification}. We consider the cases that abide the vacuum stability conditions. We use ``$\boldsymbol\times$'' to represent that the couplings are turned off to simplify the system. We gradually increase the complexity of the system from scenario 1 to 5 where more and more couplings are involved. Scenario 4 (two sample cases, for example, will be the BF~row and AF~ row in Tab.~\ref{shifting UV fixed point_1}) and scenario 5 (one sample case will be the AF~row in Tab.~\ref{shifting UV fixed point_3}) possess all the couplings involved in our Pati-Salam model.
 The value $0$ denotes a zero value solution at the fixed point. There are two distinct cases in which a specific coupling can be zero at the UV fixed point: the coupling can be asymptotically free  or can vanish at all scales. For example,  $y_F$ is asymptotically free and therefore it leads to interesting physics in the IR while $\lambda_3$, $\lambda_{R\Phi_2}$ and $\lambda_{R\Phi_3}$ can be set to zero at all energies, with the current approximations. This is what is assumed in the last row of Tab.~\ref{classification} to simplify the analysis. 
\begin{table}[t!]
\centering
  \begin{tabular}{|| l | l | l | l | l | l | l | l | l | l | l | l | l | l ||}
    	\hline
    &	$\lambda_1$ & $\lambda_2$ & $\lambda_3$ & $\lambda_4$ & $\lambda_{R\Phi_1}$ & $\lambda_{R\Phi_2}$ & $\lambda_{R\Phi_3}$ & $\lambda_{R1}$ & $\lambda_{R2}$ & $y$ & $y_c$ & $y_\nu$ & $y_F$\\ \hline
   1 & Irev & $\boldsymbol\times$ & $\boldsymbol\times$ & $\boldsymbol\times$ & Rev & Rev &$\boldsymbol\times$ & Irev & Rev & Irev & Irev & Irev & 0\\ \hline
    $2$ &Rev & $\boldsymbol\times$ & $\boldsymbol\times$ & Irev & Irev & Rev & $\boldsymbol\times$ & Irev & Rev & Irev & Irev & Irev & 0\\ \hline
    	3 &Irev & Rev & $\boldsymbol\times$ & $\boldsymbol\times$ & Rev & Irev & $\boldsymbol\times$ & Irev & Rev & Irev & Irev & Irev & 0\\ \hline 
    	4&Irev & Rev & 0 & Irev & Irev & 0 & 0 & Irev & Rev & Irev & Irev & Irev & 0\\ \hline
    	    	5&Irev & Rev & 0 & Irev & Irev & 0 & 0 & Irev & Irev & Irev & Irev & Irev & Irev\\ \hline
\end{tabular}
\caption{\small Classifications of the UV fixed point solutions of the couplings with relevant (Rev) and irrelevant (Irev) characteristics are listed. 
The symbol ``$\boldsymbol\times$'' denotes the corresponding coupling is turned off for simplification. From scenario 1 to 5, the complication of the scenario is gradually increased. $y_F=0$ is due to the asymptotically free solution we choose while $\lambda_3$, $\lambda_{R\Phi_2}$ and $\lambda_{R\Phi_3}$ are chosen to be zero at all scale for simplification.}
\label{classification}
\end{table}

We employed two approaches to determine the RG flow of the system: the IR to UV approach and the UV to IR approach. In the IR to UV approach, the RG flow of the irrelevant couplings is constrained on certain trajectories, the separatrices.\footnote{A separatrix is the globally defined trajectory dividing the RG flow into distinct physical regions.} Thus, we can solve the set of equations $\beta_i=0$ ($i$ corresponding to all the irrelevant couplings) and solve for  all the irrelevant couplings as function of the relevant couplings. The IR initial conditions of the relevant couplings are compatible with the phenomenological constraints while preserving  UV safety.  For the UV to IR approach one simply starts from the UV fixed point and attempts to run towards the IR. Here we use the fact that the gauge couplings have RG functions that are sufficiently decoupled from the other couplings. Thus, we can run the remaining couplings along the determined gauge coupling RG trajectories.  
\begin{figure}[t!]
\centering
\subfigure[$g_i-\mu$]{
\begin{minipage}{7cm}
\centering
\includegraphics[width=0.8\columnwidth]{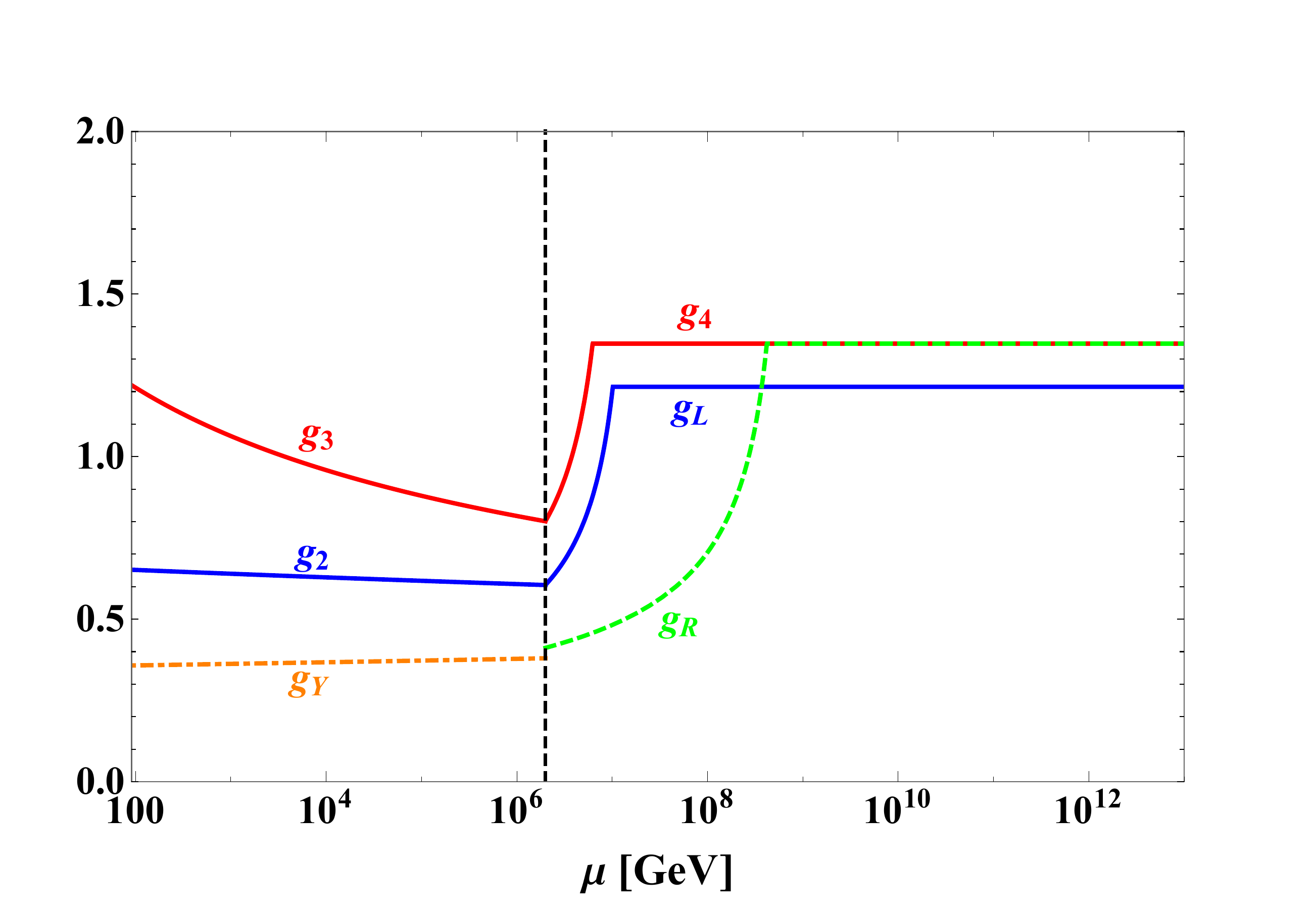}\hspace{0.06\columnwidth}
\end{minipage}
}
\subfigure[$y-\mu$]{
\begin{minipage}{7cm}
\centering
\includegraphics[width=0.8\columnwidth]{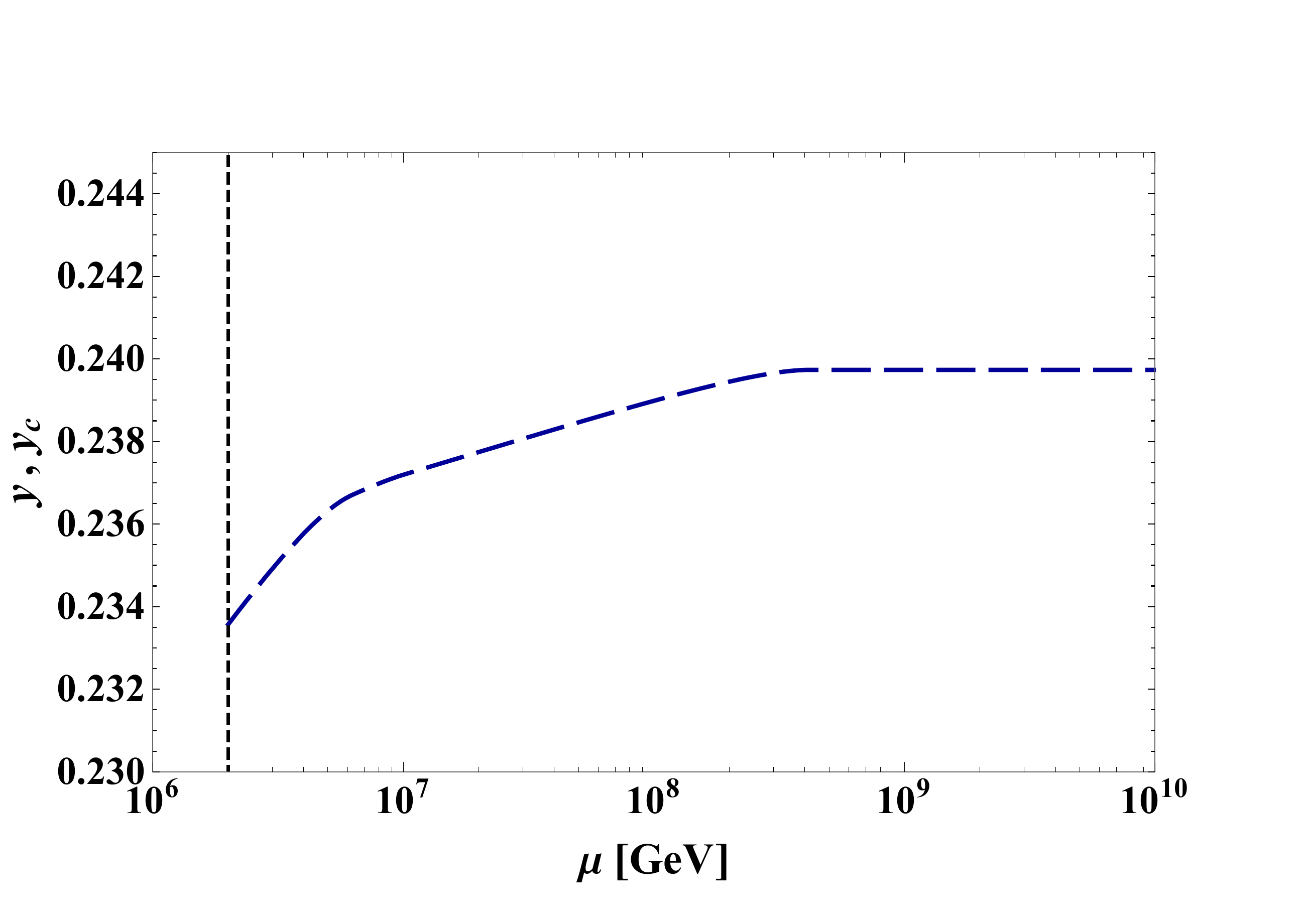}\hspace{0.06\columnwidth}
\end{minipage}
}
\subfigure[$y_\nu-\mu$]{
\begin{minipage}{7cm}
\centering
\includegraphics[width=0.8\columnwidth]{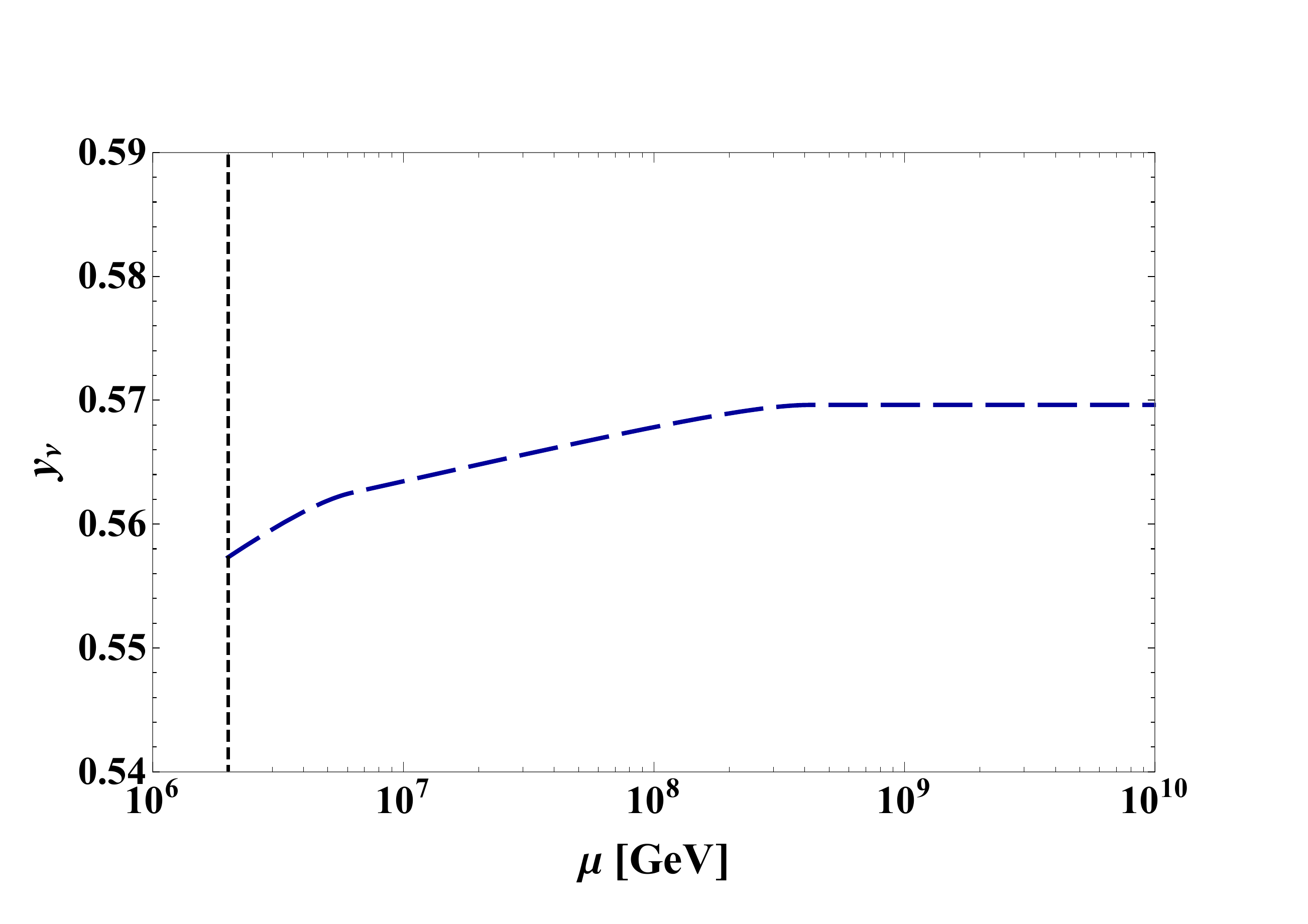}\hspace{0.06\columnwidth}
\end{minipage}
}
\subfigure[$y_F-\mu$]{
\begin{minipage}{7cm}
\centering
\includegraphics[width=0.8\columnwidth]{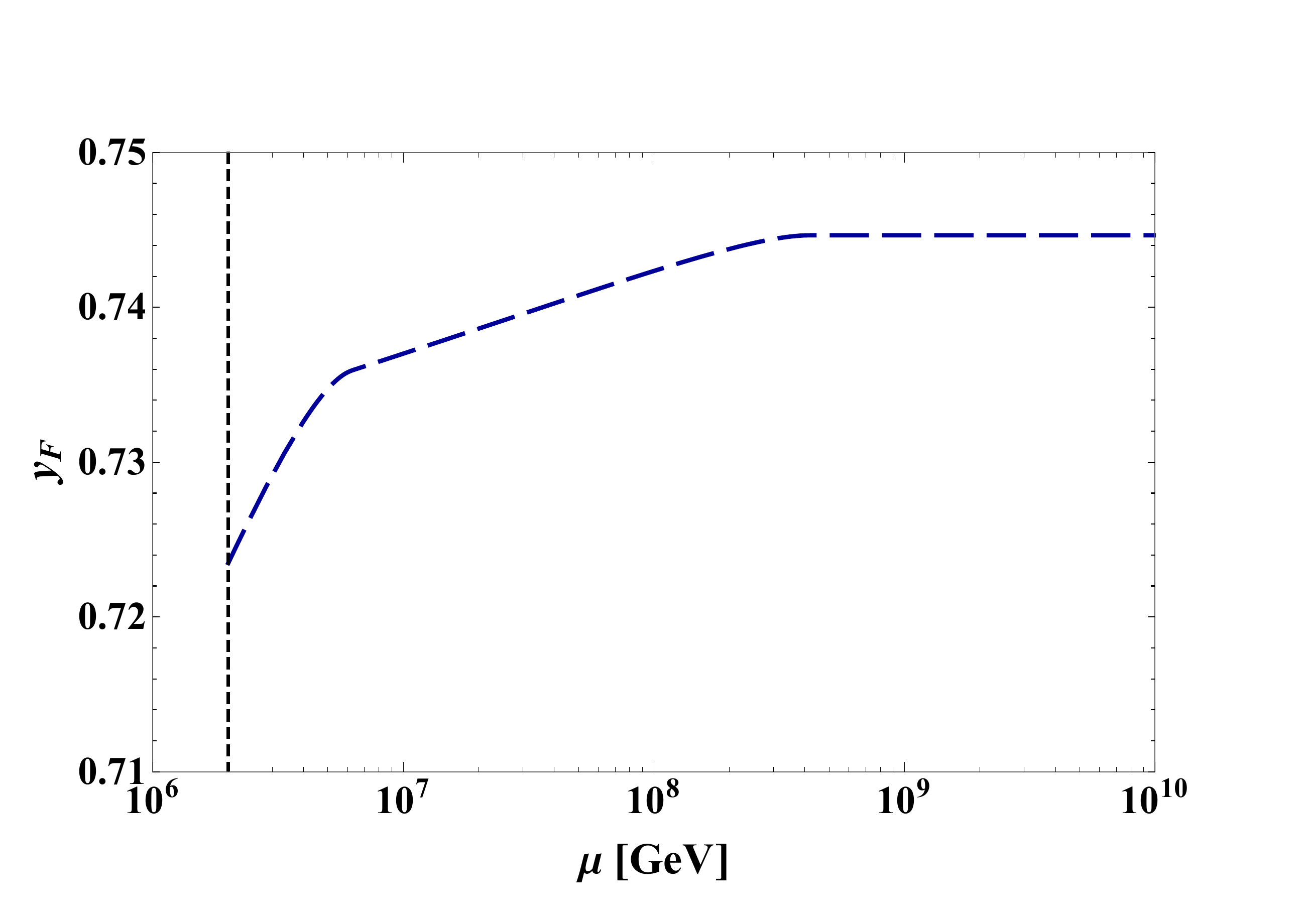}\hspace{0.06\columnwidth}
\end{minipage}
}
\caption{\small RG running of the gauge and Yukawa couplings by using the UV to IR approach. We have chosen $N_{F2}=40,\,N_{F3}=130,\,N_{F4}=130$. We have used the matching conditions at IR (see Eq.~\eqref{matching_1} and Eq.~\eqref{matching_2}) to set the initial conditions of $g_L,\,g_R,\,g_4$ at IR. For simplification, we have assumed that the vector-like fermions under gauge different symmetry groups are exactly introduced at  the Pati-Salam breaking scale, $v_R=2000\,\rm{TeV}$, marked by a vertical dashed line.}
\label{Running_Couplings_1}
\end{figure}

\begin{figure}[htb]
\centering
\subfigure[$\lambda_1-\mu$]{
\begin{minipage}{7cm}
\centering
\includegraphics[width=0.8\columnwidth]{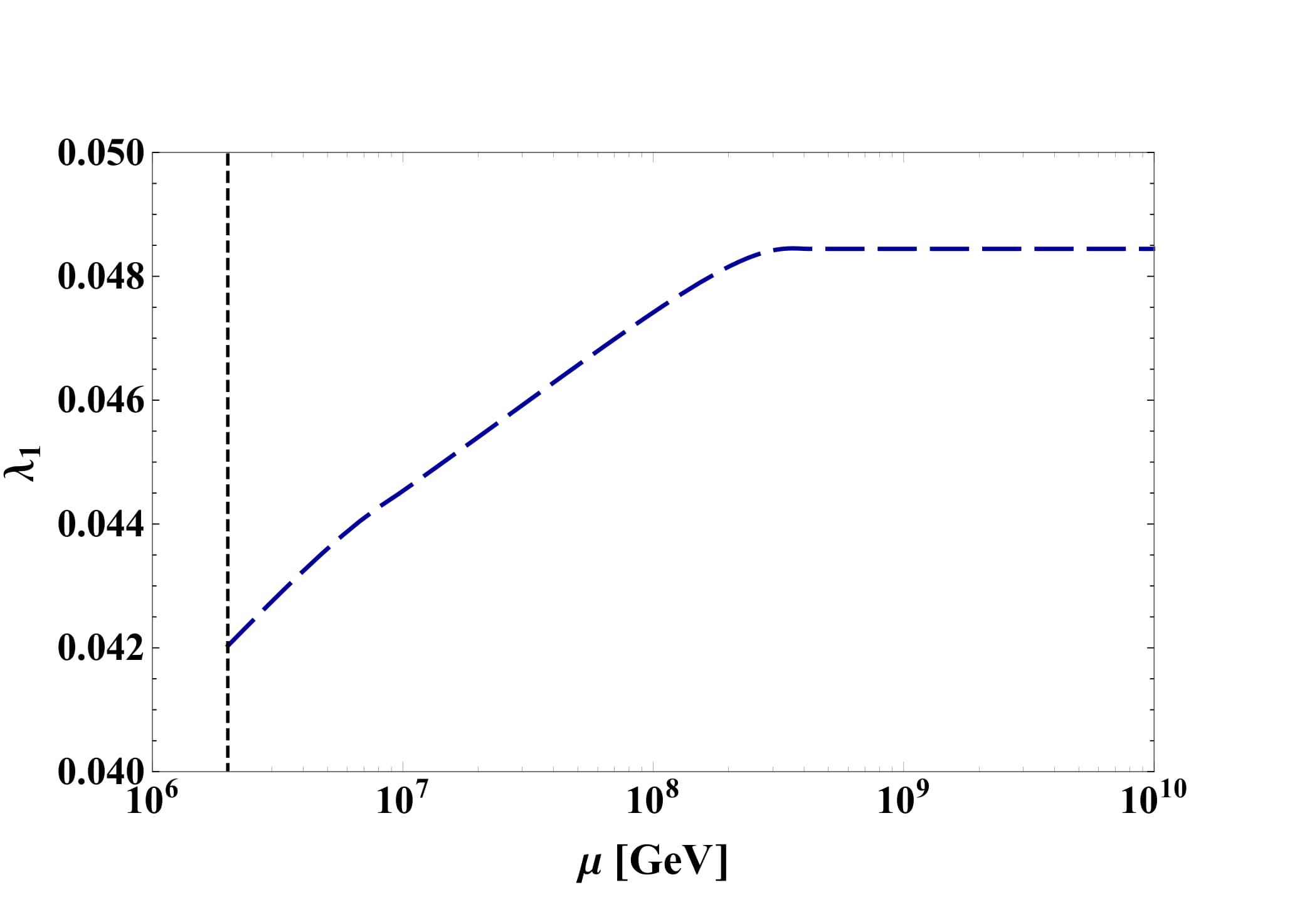}\hspace{0.05\columnwidth}
\end{minipage}
}
\subfigure[$\lambda_2-\mu$]{
\begin{minipage}{7cm}
\centering
\includegraphics[width=0.8\columnwidth]{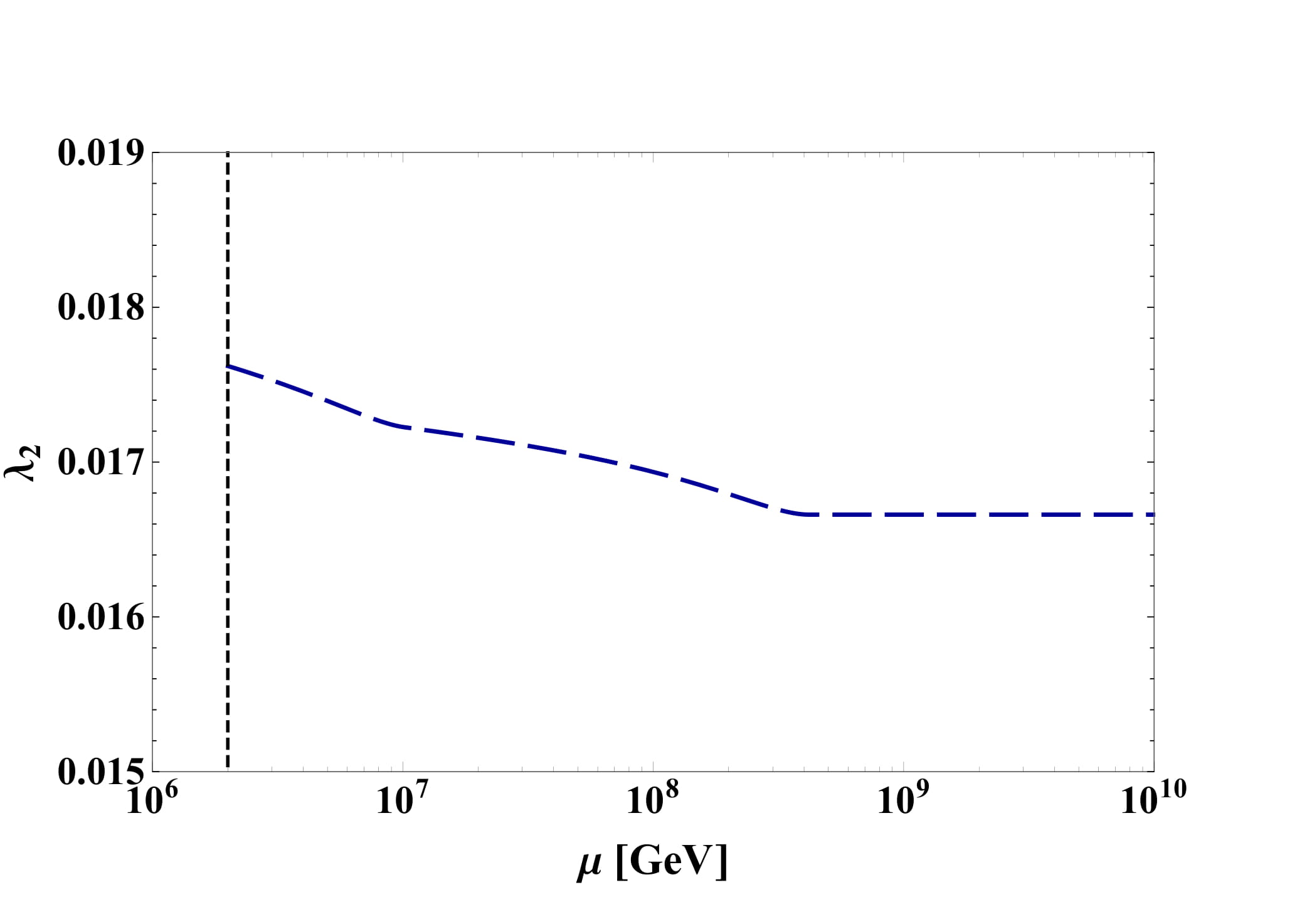}\hspace{0.05\columnwidth}
\end{minipage}
}
\subfigure[$\lambda_4-\mu$]{
\begin{minipage}{7cm}
\centering
\includegraphics[width=0.8\columnwidth]{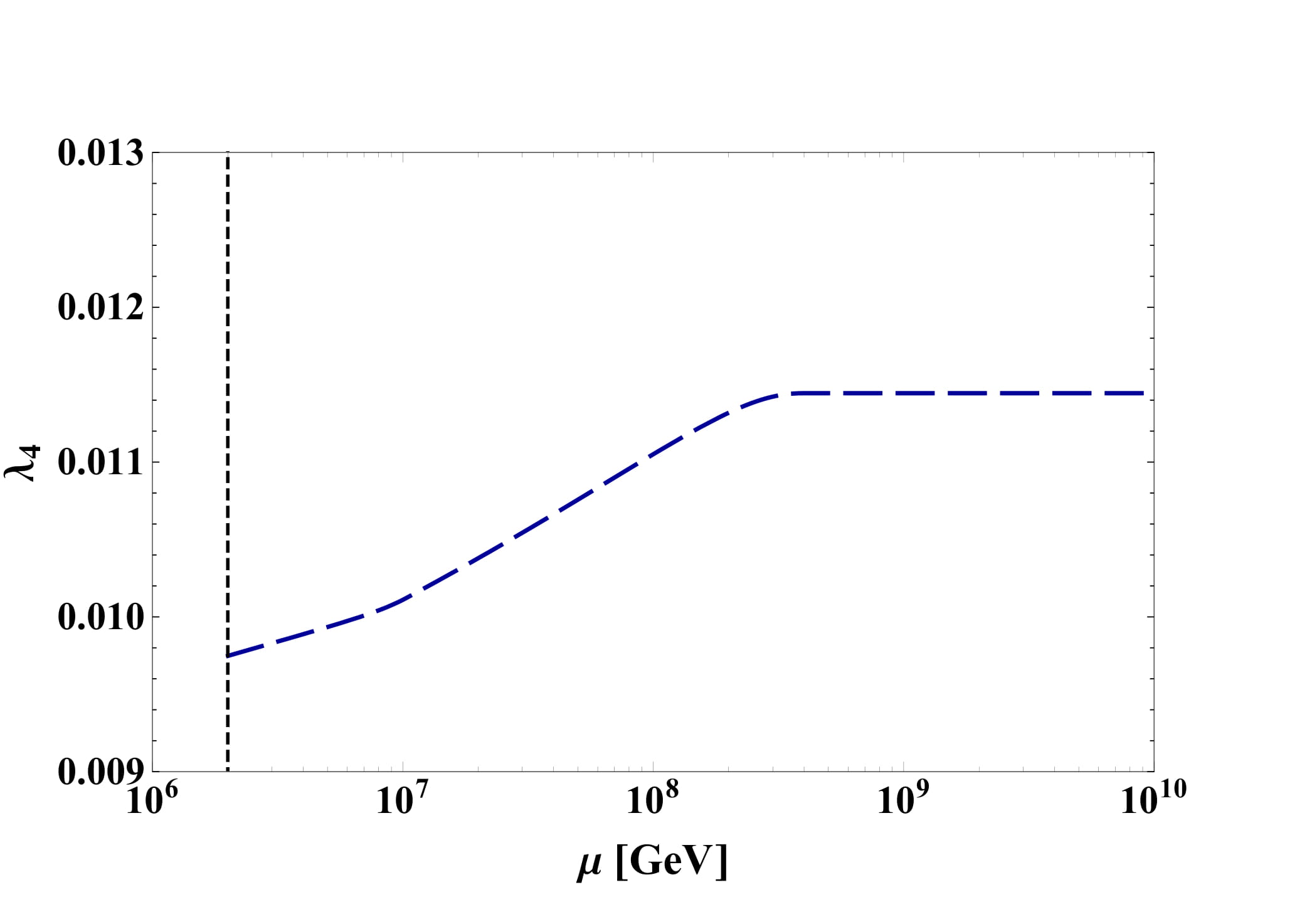}\hspace{0.05\columnwidth}
\end{minipage}
}
\subfigure[$\lambda_{R\phi_1}-\mu$]{
\begin{minipage}{7cm}
\centering
\includegraphics[width=0.8\columnwidth]{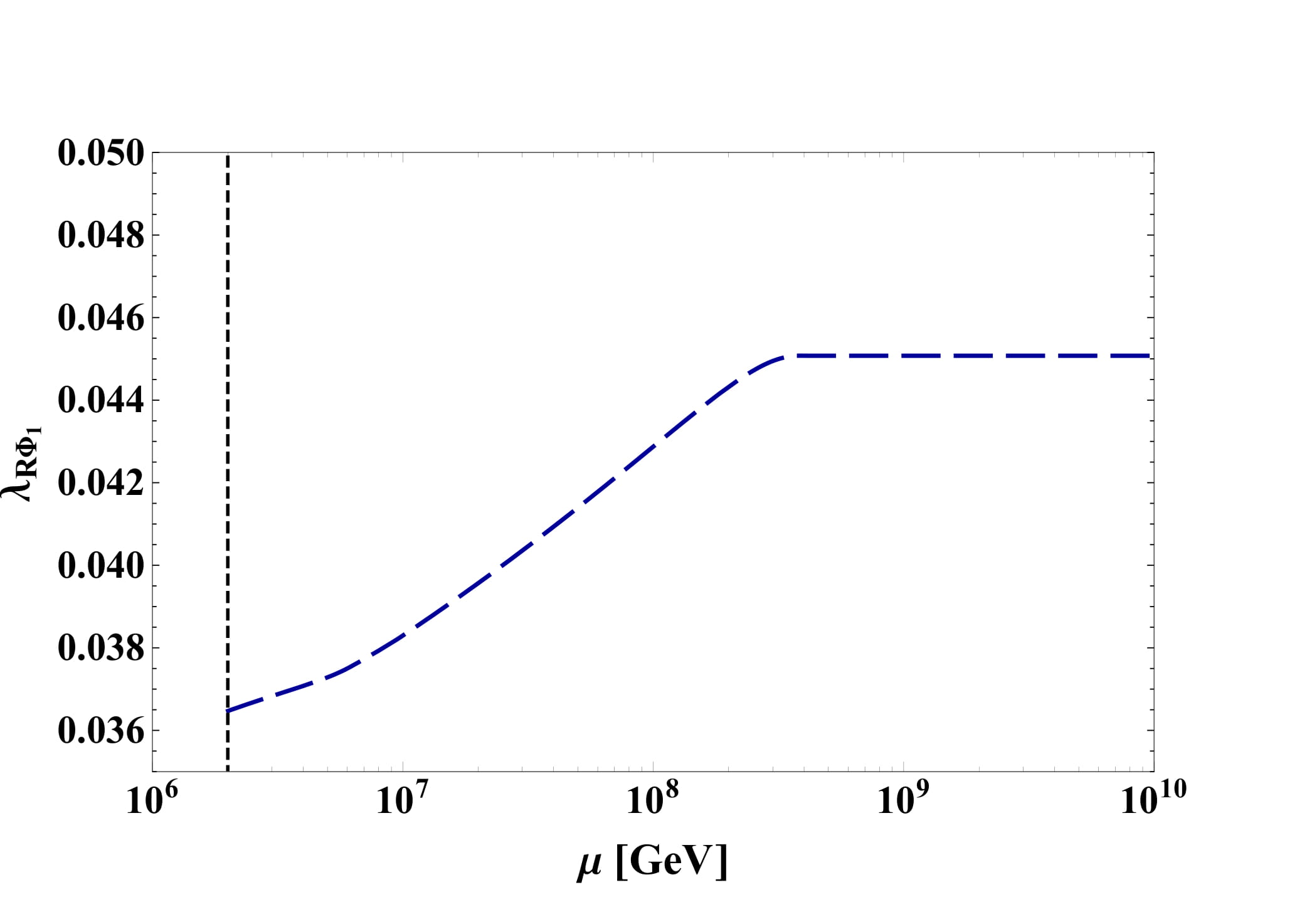}\hspace{0.05\columnwidth}
\end{minipage}
}
\subfigure[$\lambda_{R1}-\mu$]{
\begin{minipage}{7cm}
\centering
\includegraphics[width=0.8\columnwidth]{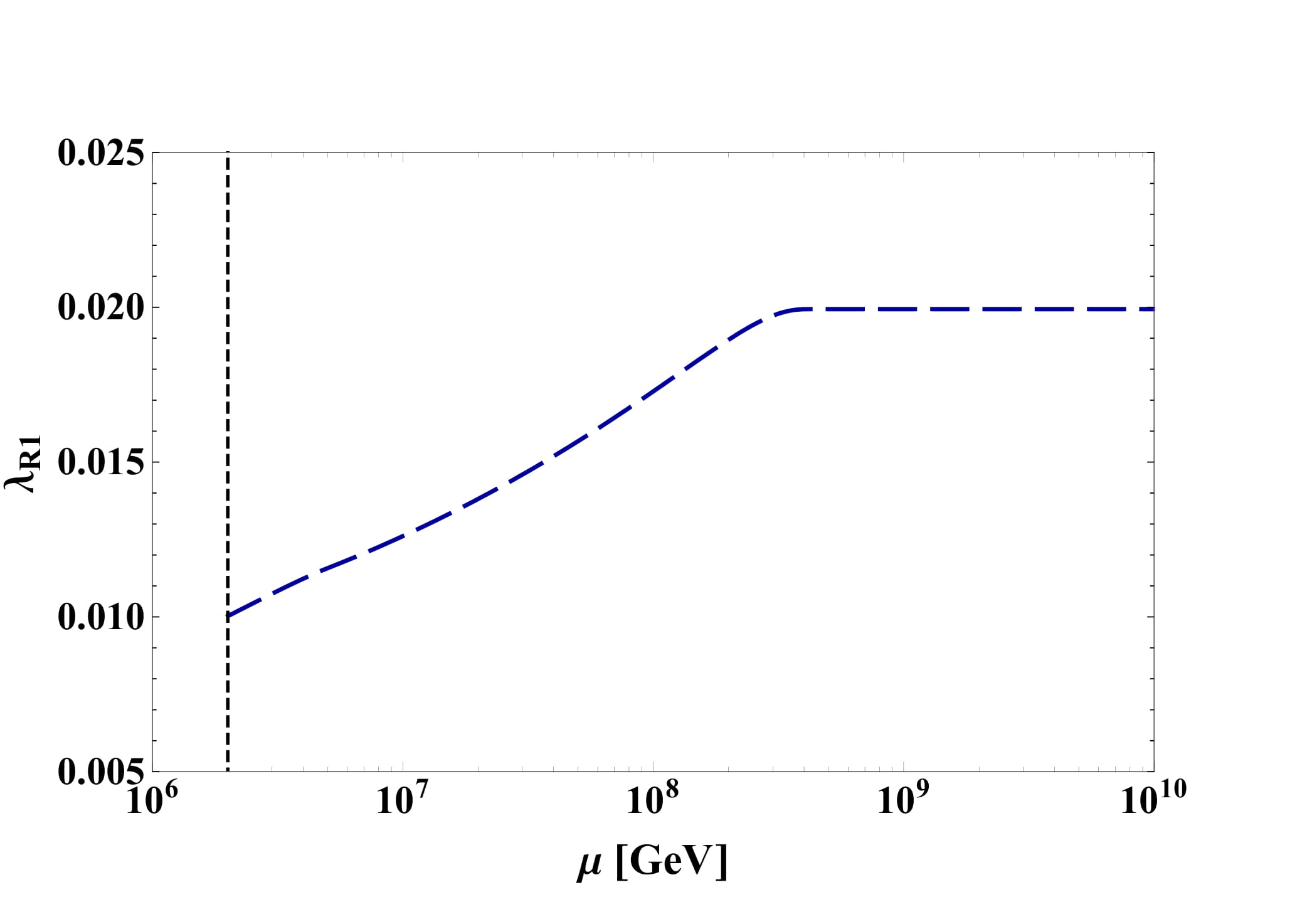}\hspace{0.05\columnwidth}
\end{minipage}
}
\subfigure[$\lambda_{R2}-\mu$]{
\begin{minipage}{7cm}
\centering
\includegraphics[width=0.8\columnwidth]{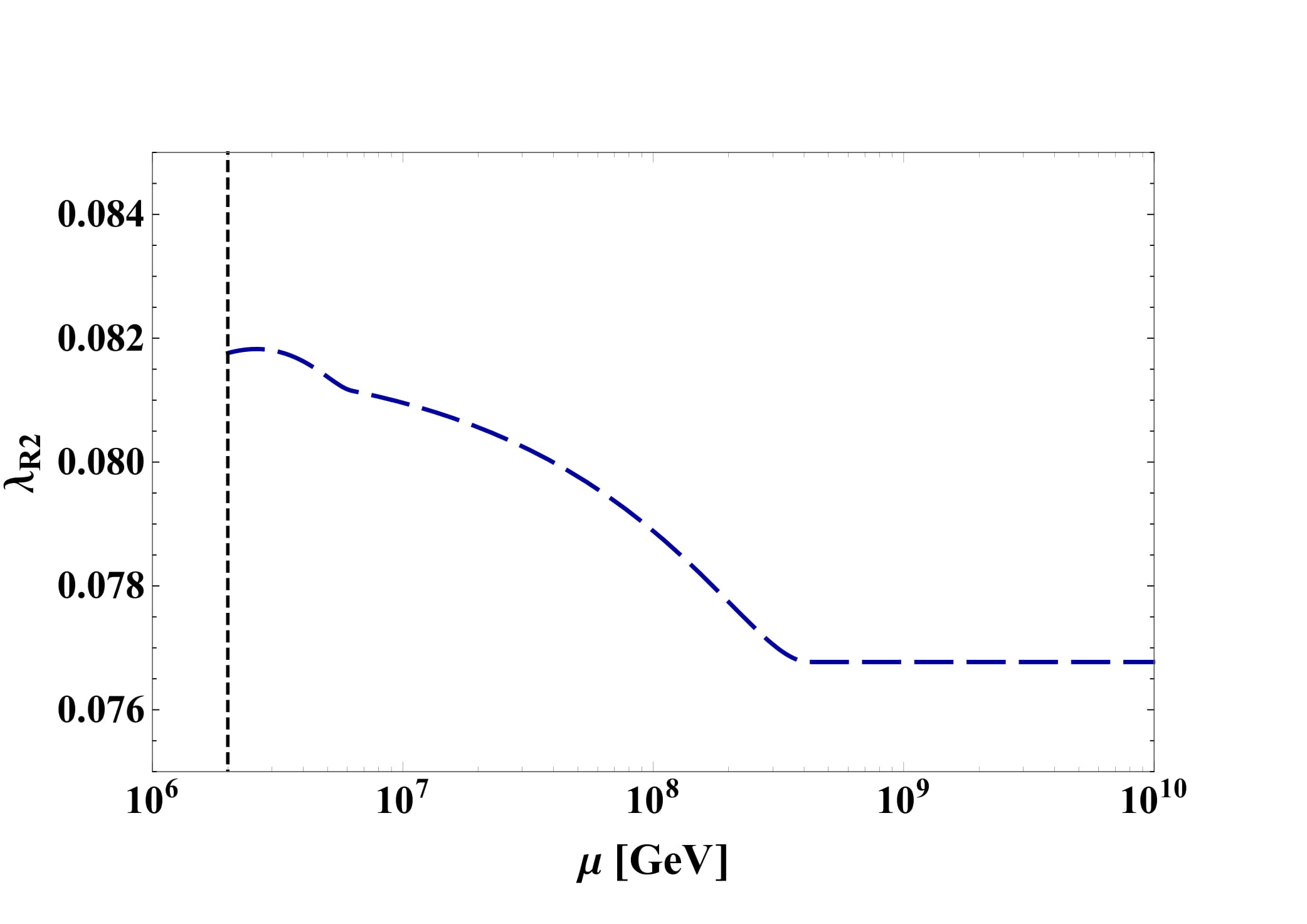}\hspace{0.05\columnwidth}
\end{minipage}
}
\caption{\small RG running of the scalar quartic couplings using the UV to IR approach for $N_{F2}=40,\,N_{F3}=130,\,N_{F4}=130$. All the vector-like fermions  appear (dashed line) at the symmetry breaking scale of the Pati-Salam group, which is around  $2000\,\rm{TeV}$.}
\label{Running_Couplings_2}
\end{figure}

We report our results in Figs.~\ref{Running_Couplings_1} and \ref{Running_Couplings_2} where we show the running of the gauge, Yukawa and scalar couplings by using the UV to IR approach for  ($N_{F_{2L}}=40,\,N_{F_{2R}}=130,\,N_{F_4}=130$). The corresponding UV fixed point solution is the one shown in the AF~row of Tab.~\ref{shifting UV fixed point_3}. As mentioned above the RG flows of the gauge couplings are determined once the IR conditions are given. The IR initial conditions for $g_L$, $g_R$ and $g_4$ are obtained by using the matching conditions of Eq.~\eqref{matching_1} and Eq.~\eqref{matching_2} and the SM couplings are running from the EW scale to the Pati-Salam symmetry breaking scale. For simplicity, the vector-like fermions masses are taken to be the Pati-Salam symmetry breaking scale $v_R=2000\,\rm{TeV}$. From Figs.~\ref{Running_Couplings_1} and \ref{Running_Couplings_2}, it is  clear that all couplings (i.e.~gauge, Yukawa and scalar quartic) achieve a safe UV fixed point. The transition scale, above which,  the UV fixed point is reached is about $0.5\times10^9\,\rm{GeV}$ for all the couplings. Note that we could shift this transition scale significantly by increasing (the scale will decrease) or decreasing (the scale will increase) the number of vector-like fermions.

\section{Matching the Standard Model}\label{Sec4}
We now consider gluing the ultraviolet safe theory to the SM couplings at low energies, which is an important test in order to render our high energy safe extension phenomenologically viable. We start with the scalar component of the theory.

\subsection{Scalar Sector}
After Pati-Salam symmetry breaking, the scalar bi-doublet should match the conventional two Higgs doublet model which is defined by the Lagrangian:
\begin{equation}
\begin{split}
V_H&=m_{11}^2\Phi_1^\dagger\Phi_1+m_{22}^2\Phi_2^\dagger\Phi_2-\left(m_{12}^2\Phi_1^\dagger\Phi_2+\rm{H.c.}\right)\\
&+\bar{\lambda}_1\left(\Phi_1^\dagger\Phi_1\right)^2+\bar{\lambda}_2\left(\Phi_2^\dagger\Phi_2\right)^2+\bar{\lambda}_3\left(\Phi_1^\dagger\Phi_1\right)\left(\Phi_2^\dagger\Phi_2\right)+\bar{\lambda}_4\left(\Phi_1^\dagger\Phi_2\right)\left(\Phi_2^\dagger\Phi_1\right)\\
&+\left[\frac{1}{2}\bar{\lambda}_5\left(\Phi_1^\dagger\Phi_2\right)^2+\bar{\lambda}_6\left(\Phi_1^\dagger\Phi_1\right)\left(\Phi_1^\dagger\Phi_2\right)+\bar{\lambda}_7\left(\Phi_2^\dagger\Phi_2\right)\left(\Phi_1^\dagger\Phi_2\right)+\rm{H.c.}\right]\,.\label{two Higgs doublet}
\end{split}
\end{equation}
Comparing  \eqref{two Higgs doublet} with \eqref{potential}, we find:
\begin{equation}
\begin{split}
&\bar{\lambda}_1=\lambda_1,\quad\bar{\lambda}_2=\lambda_1,\quad\bar{\lambda}_3=2\lambda_1,\quad\bar{\lambda}_4=4\left(-2\lambda_2+\lambda_4\right)\\
&\bar{\lambda}_5=4\lambda_2,\quad\bar{\lambda}_6=-\lambda_3,\quad\bar{\lambda}_7=\lambda_3\label{quartic_matching}\,.
\end{split}
\end{equation}
When a set of $\left(N_{F2},\,N_{F3},\,N_{F4}\right)$ is given and a Pati-Salam symmetry breaking pattern is chosen, by using the RG running from a specific UV fixed point, we could predict the coupling values at the Pati-Salam symmetry breaking scale. After implementing the matching conditions of Eq.~\eqref{quartic_matching} these couplings become our new initial values so that by employing the two Higgs doublet RG beta functions \cite{Branco:2011iw}, we could obtain the coupling values at the electroweak scale.

At this point we turn our attention to the mass matrix (neutral scalar fields) of the two Higgs doublet model:
\begin{equation}
M^2_{\rm{neutral}}=\left[
\begin{array}{cc}
\frac{m_{12}^2 v_2}{v_1}+2 \bar{\lambda}_1 v_1^2 & -m_{12}^2+\left(\bar{\lambda} _3+\bar{\lambda}_4+\bar{\lambda} _5\right) v_1 v_2 \\
-m_{12}^2+\left(\bar{\lambda}_3+\bar{\lambda} _4+\bar{\lambda}_5\right) v_1 v_2 & \frac{m_{12}^2 v_2}{v_1}+2 \bar{\lambda} _2 v_2^2 \\
\end{array}
\right]\,.\label{Mass_two_doublet}
\end{equation}
Note that this mass matrix is defined at the electroweak scale and to make sure that the computation is complete, we further included the higher order corrections for the mass matrix (e.g. the RG improved mass matrix), which for simplicity are not shown explicitly in Eq.~\eqref{Mass_two_doublet}. By using the coupling values obtained previously at the electroweak scale, we determine the mass eigenvalues. The important phenomenological constraints are: both eigenvalues of the mass matrix should be positive and the lighter one should be close to the $125\,\rm{GeV}$ value of the observed Higgs mass. It can be shown that by choosing $N_{F2}=32,\,N_{F3}=108,\,N_{F4}=56$, we obtain the following coupling values at around the electroweak scale: 
\begin{equation}
\bar{\lambda}_1=0.222,~\bar{\lambda}_2=0.222,~\bar{\lambda}_3=0.250,~\bar{\lambda}_4=-0.380,~\bar{\lambda}_5=0.260,~y=0.614\,.\label{couplingvalue_EW}
\end{equation}
It is interesting to discuss two cases in the following: the one for which we set $m_{12}=0$ and the other for $m_{12}\neq 0$. For the case with $m_{12}=0$, we obtain two neutral scalar masses, one with $\sim92\,\rm{GeV}$ (lighter Higgs) and an heavier one of $\sim123\,\rm{GeV}$. We stress that the choice $\left(N_{F2}=32,\,N_{F3}=108,\,N_{F4}=56\right)$ is among the ones in which one can achieve the heaviest Higgs mass. We find intriguing that to push the lighter Higgs to be closer to the observed Higgs mass requires the Pati-Salam symmetry breaking scale not to be too far away from $10^4\,\rm{TeV}$. Overall,  for  $m_{12}=0$, both mass eigenvalues are phenomenologically too light. For the case with $m_{12}\neq0$, we find both mass eigenvalues to increase when increasing $m_{12}$. In particular the light Higgs for the above choice of $N_F$ seems to converge towards the value of $120\,\rm{GeV}$ (almost not changing after $m_{12}>150\,\rm{GeV}$) while the heavier Higgs mass keeps increasing with $m_{12}$. When choosing $m_{12}=150\,\rm{GeV}$, the light Higgs and the heavy Higgs masses are respectively $120\,\rm{GeV}$ and $264\,\rm{GeV}$. We find that also in this case, there is a constraint on the Pati-Salam symmetry breaking scale. Slightly different from the case where $m_{12}=0$, it now requires the Pati-Salam symmetry breaking scale to be larger than $10^4\,\rm{TeV}$ in order to yield a phenomenologically viable Higgs mass.

We also found alternative RG flow solutions leading to two light Higgs (we briefly list the result here). For $N_{F2}=36,\,N_{F3}=109,\,N_{F4}=118$, we could have one light Higgs at $22.6\,\rm{GeV}$ and the heavy one at $125\,\rm{GeV}$. We point to recent studies in the detection of light scalars~\cite{Chang:2017ynj}\cite{Dupuis:2016fda}.
\begin{figure}[t!]
\centering
\includegraphics[width=0.4\columnwidth]{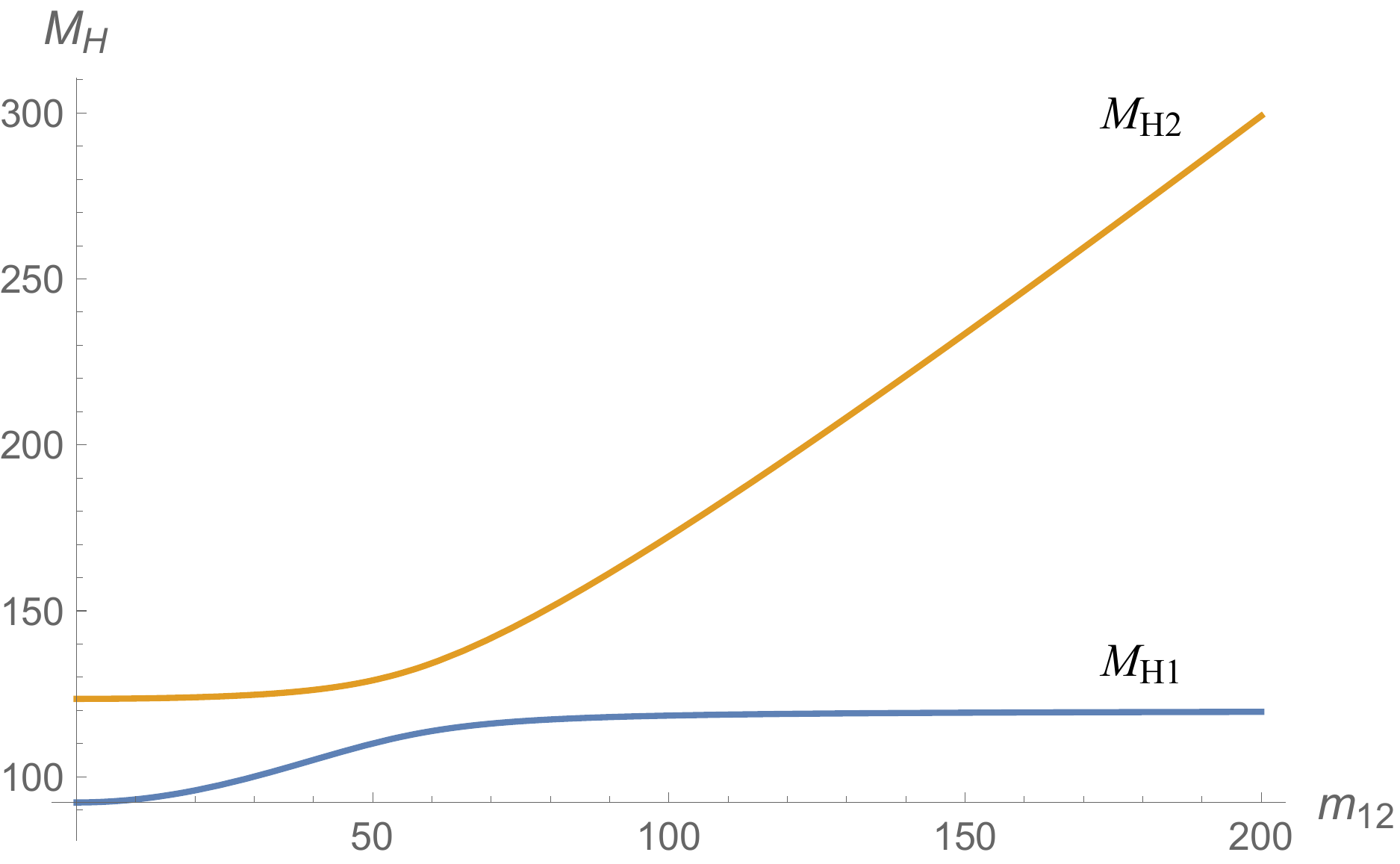}\hspace{0.04\columnwidth}
\caption{\small In this figure, we show two mass eigenvalues of the CP-even neutral Higgs mass matrix as a function of $m_{12}$.  $M_{H1}$ denotes the lighter Higgs while $M_{H2}$ denotes the heavier one.}
\label{unification}
\end{figure}

\subsection{Yukawa Sector}
The top Yukawa mass term by using Eq.~\eqref{spectrum1} at the electroweak scale is given by :
\begin{equation}
	m_{\rm{top}}  =  (y \sin \beta + y_c \cos\beta) v=\sqrt{2}yv\,,
\end{equation} 
where we have implemented the CP symmetry leading to $y=y_c$ and $\tan\beta=1$. Thus it is clear that to obtain the corret top quark mass at the electroweak scale, y should be $\sim\frac{0.93}{\sqrt{2}}\sim0.66$ which is smaller than the conventional SM top Yukawa coupling value. From Eq.~\eqref{couplingvalue_EW}, we obtain $y=0.614$ which is  close to the required value.

To obtain the correct bottom quark mass, we have discussed previously that it requires to introduce a new $10-\rm{dim}$ vector-like fermion $F\sim (10, 1, 1)$ with mass $M_F$ to trigger the bottom-top mass splitting. The colour triplet $B$ component of $F\sim (10, 1, 1)$ obtains the mass after Pati-Salam symmetry breaking: $m_B\equiv y_F\, v_R/\sqrt{2}\sim7000\,\rm{TeV}$, where we have used the same set of $N_{F2}=32,\,N_{F3}=108,\,N_{F4}=56$ above providing $v_R\sim10000\,\rm{TeV}$ and $y_F=1$ at the Pati-Salam Symmetry breaking scale. By using Eq.~\eqref{bottom_tau_mass}, it requires $M_F=\frac{\sqrt{2}m_Bm_b}{m_t}=239.2\,\rm{TeV}$ to obtain the correct bottom quark mass. Since $M_F$ is a free parameter in our theory, we do have the freedom to choose the desired value.

\section{Conclusions}
\label{Sec5}
Models in which scalar degrees of freedom are fundamental a l\' a Wilson \cite{Wilson:1971bg,Wilson:1971dh} require the presence of scale invariance at short distances \cite{Antipin:2013exa,Pelaggi:2017wzr,Shaposhnikov:2018jag}. A complete safe or free theory can therefore support elementary scalars as fundamental fields.  Fundamentality and naturality are complementary concepts. Short distance scale invariance implies fundamentality while (near) long distance conformality and/or  controllably broken symmetries help with naturality \cite{Antipin:2013exa,Pelaggi:2017wzr}.  A coherent search of safe extensions of the SM has only recently begun.  In this work we have constructed a realistic safe extension of the SM in which we add vector-like fermions to the time-honored Pati-Salam framework.  Recent progress in the large-$N$ safe dynamics of gauge-Yukawa theories has proven instrumental for the success of the project. In particular we have shown that the gauge, scalar quartic and Yukawa couplings achieve an interacting ultraviolet fixed point below the Planck scale. The minimal model is a relevant example of a Standard Model extension in which unification of {\it all} type of couplings occurs because of a dynamical principle, i.e. the presence of an ultraviolet fixed point. Most importantly, we are able to show that starting from specific UV fixed points, some of the RG flows can match both the SM Higgs mass and Yukawa couplings (top and bottom) which implies a truly UV completion of the Standard Model. It is also intriguing that, in this minimal model, the Pati-Salam symmetry breaking scale is close to $10^4\,\rm{TeV}$ to yield a physically acceptable Higgs mass. There are several aspects that deserve further investigation from a more in depth phenomenological study of the quark and lepton flavour sector to baryogenesis. 

\section{Acknowledgments}
The work is partially supported by the Danish National Research Foundation under the grant DNRF:90 and the Natural Sciences and Engineering Research Council of Canada (NSERC). E.~Molinaro thanks the Department of Physics and Astronomy of Aarhus University for the hospitality during the completion of this paper. Z.W.~Wang thanks Robert Mann, Tom Steele, Heidi Rzehak, Chen Zhang and Jing Ren for very helpful suggestions.
\pagebreak

\appendix 
\section{One-Loop RG equations of the Pati-Salam model}\label{RG_all}
\label{App1}
\subsection{Gauge couplings}
		 \begin{align}
						(4\pi)^2\beta_{g_{L}} = 
					&- 3g_{L}^{3}\,,\\
					(4\pi)^2\beta_{g_{4}} = 
					&- 9 g_{4}^{3}\,,\\
					(4\pi)^2\beta_{g_{R}} = 
					&- \frac{7}{3} g_{R}^{3}\,.
					\end{align}
\subsection{Quartic coupling}
		\begin{align}
(4\pi)^2\beta_{\lambda_1}=
				&\lambda_1\left(32y_c^2+128 \lambda_1-128\lambda_2+64\lambda_4- 9g_{L}^{2}- 9g_{R}^{2}+32y^2\right)+512\lambda_2^2  \nonumber\\
				&+ \frac{9}{32}\left(g_{L}^{4} + g_{R}^{4}\right) + \frac{3}{16}g_{L}^{2}g_{R}^{2} + 32 \lambda_{R\Phi 1} \lambda_{R\Phi 3}  +32 \lambda_{R\Phi 1}^{2}+16                  \lambda_{R\Phi 3}^{2}\nonumber\\
				&+64\lambda_4^2- 8y^4-8y_c^4,\,\\
(4\pi)^2\beta_{\lambda_2}=
                   &\lambda_2\left(32y^2-9g_L^2-9g_R^2+96\lambda_1+192\lambda_4-384\lambda_2+32y_c^2\right)\nonumber\\
				&-4y_c^2y^2+48\lambda_3^2+32\lambda_{R\Phi2}^2\,,\\
(4\pi)^2\beta_{\lambda_3}=
                  &\lambda_3\left(32y_c^2-9g_L^2-9g_R^2+192\lambda_1+192\lambda_4+32y^2\right)-8y^3y_c-8yy_c^3\nonumber\\
                  &+64\lambda_{R\phi1}\lambda_{R\phi2}+32\lambda_{R\phi2}\lambda_{R\phi3}\\
(4\pi)^2\beta_{\lambda_4}=
				&\lambda_4\left(32y^2-9g_L^2-9g_R^2+96\lambda_1+128\lambda_2+64\lambda_4+32y_c^2\right)+\frac{3}{8}g_L^2g_R^2\nonumber\\
				&+192\lambda_3^2+128\lambda_{R\Phi2}^2-8\lambda_{R\Phi3}^2-24y^2y_c^2+4y_c^4+4y^4\,,\\
(4\pi)^2\beta_{\lambda_{R1}}=&\lambda_{R1}\left(192 \lambda_{R1} - 9 g_{R}^{2}- \frac{45}{2} g_{4}^{2}+192 \lambda_{R2} +20y_{F}^2 + 8 y_{\nu}^2\right)+ 48 \lambda_{R2}^{2}\nonumber \\
				&+ \frac{9}{32}g_{R}^{4} + \frac{27}{128}g_{4}^{4}+ \frac{27}{32} g_{R}^{2}g_{4}^{2}+ 16 \lambda_{R\Phi 1}^{2}  + 16 \lambda_{R\Phi 1} \lambda_{R\Phi 3}+64\lambda_{R\Phi2}^2\nonumber\\
				&- \frac{1 }{2} y_{F}^4- 2 y_\nu^4\,,\\
(4\pi)^2\beta_{\lambda_{R2}}=&\lambda_{R2}\left(- 9g_{R}^{2}- \frac{45}{2} g_{4}^{2} +20 y_{F}^2+ 8 y_{\nu}^2+96\lambda_{R1}\right)-\frac{9}{16}g_4^2g_R^2 +\frac{9}{16}g_4^4\nonumber\\
				&-3y_F^4+8\lambda_{R\Phi3}^2  \,,\\
(4\pi)^2\beta_{\lambda_{R\Phi 1}}=&\lambda_{R\Phi1}\bigg(16y_c^2 +16y^2 - \frac{9}{2} g_{L}^{2} - \frac{45}{4} g_{4}^{2} - 9 g_{R}^{2}+80 \lambda_1-64\lambda_2+32\lambda_4+ 144 \lambda_{R1}\nonumber\\
				&+96 \lambda_{R2}+ 10 y_{F}^2 + 4 y_{\nu}^2\bigg)+32 \lambda_1 \lambda_{R\Phi 3} + 16 \lambda_{R\Phi 1}^{2} + 64 \lambda_{R\Phi 2}^{2} +8 \lambda_{R\Phi 3}^{2}+ \frac{9 g_{R}^{4}}{16}\nonumber\\
				&  + 64 \lambda_{R1} \lambda_{R\Phi 3} - 64 \lambda_{2} \lambda_{R\Phi 3}+ 32 \lambda_{4} \lambda_{R\Phi 3}+ 16 \lambda_{R2} \lambda_{R\Phi 3}- 4 y_c^2    y_{\nu}^2-10y^2y_F^2\,, \\\nonumber
(4\pi)^2\beta_{\lambda_{R\Phi 2}}=&\lambda_{R\Phi2}\bigg(16y^2+16y_c^2 - \frac{9}{2} g_{L}^{2}- 9 g_{R}^{2}- \frac{45}{4} g_{4}^{2}+16 \lambda_1   +64\lambda_2+64\lambda_4\nonumber\\
				& + 144 \lambda_{R1}+ 96 \lambda_{R2}+ 10 y_{F}^2+32\lambda_{R\Phi 1}+16\lambda_{R\Phi 3}+ 4y_{\nu}^2\bigg) \nonumber\\
				&  +48\lambda_3\lambda_{R\Phi1}+24\lambda_3\lambda_{R\Phi3} -2 yy_c y_{\nu}^2   - 5 yy_c y_{F}^2 \,,\\\nonumber
				\end{align}
				
\begin{align}				
(4\pi)^2\beta_{\lambda_{R\Phi 3}}=&\lambda_{R\Phi3}\bigg(16y^2+16y_c^2-\frac{9}{2}g_L^2-\frac{45}{4}g_4^2+16\lambda_1+64\lambda_2-32\lambda_4+16\lambda_{R1}\nonumber\\
                  &+64\lambda_{R2}+10y_F^2+32\lambda_{R\Phi1}+4y_\nu^2\bigg)+16\lambda_{R\Phi3}^2-10y_c^2y_F^2 +10y^2y_F^2\nonumber\\
                  &+4y_c^2y_\nu^2 -4y^2y_\nu^2\,.
\end{align}
			
\subsection{Yukawa couplings}
		\begin{align}
(4\pi)^2\beta_{y}=
				&- \frac{9 y}{4} g_{L}^{2} - \frac{9 y}{4} g_{R}^{2} - \frac{45 y}{4} g_{4}^{2} + 12 y^{3} + y y_{\nu}^2
				+\frac{5}{2} y y_{F}^2\,,\\
(4\pi)^2\beta_{y_c}=
				&- \frac{9 y_c}{4} g_{L}^{2} - \frac{9 y_c}{4} g_{R}^{2} - \frac{45 y_c}{4} g_{4}^{2} + 12 y_c^{3} + y_c y_{\nu}^2
				+\frac{5}{2} y_c y_{F}^2\,,\\
(4\pi)^2\beta_{y_{F}}=
				&- \frac{9 y_{F}}{4} g_{R}^{2} - \frac{153 y_{F}}{8} g_{4}^{2} + \frac{19}{2} y_{F}^{3} -  y_{F} y_{\nu}^2
				+2 y^2 y_{F}\,,\\
(4\pi)^2\beta_{y_{\nu}}=
				&- \frac{9 y_{\nu}}{4} g_{R}^{2} - \frac{45 y_{\nu}}{8} g_{4}^{2} - \frac{5}{2} y_{F}^2 y_{\nu} + 11 y_{\nu}^{3} 
				+4y^2 y_{\nu}\,. 
\end{align}


\end{document}